\documentclass[prb,amsmath,showpacs,twocolumn]{revtex4-1}
\usepackage{graphicx}
\usepackage{multirow}
\usepackage{amssymb}
\usepackage{subfigure}
\usepackage{array}
\usepackage{appendix}
\usepackage{longtable}

\begin{document}

\title{First-principles predictions of low-energy phases of multiferroic
  BiFeO$_3$}

\author{Oswaldo Di\'eguez, O.E. Gonz\'alez-V\'azquez, Jacek
  C. Wojde\l, and Jorge \'I\~niguez}

\address{Institut de Ci\`encia de Materials de Barcelona
(ICMAB-CSIC), Campus UAB, 08193 Bellaterra, Spain}

\begin{abstract}
We used first-principles methods to perform a systematic search for
potentially-stable phases of multiferroic BiFeO$_3$. We considered a
simulation cell compatible with the atomic distortions that are most common
among perovskite oxides, and found a large number of local minima of the
energy within 100~meV/f.u. of the ferroelectric ground state. We discuss the
variety of low-symmetry structures discovered, as well as the implications of
these findings as regards current experimental (e.g., on thin films displaying
{\em super-tetragonal} phases) and theoretical (on models for BiFeO$_3$'s
structural phase transitions) work on this compound.
\end{abstract}

\pacs{77.84.-s, 75.85.+t, 71.15.Mb, 61.50.Ah}

% 77. Dielectrics, piezoelectrics, and ferroelectrics and their properties
% 77.80.-e Ferroelectricity and antiferroelectricity
% 77.84.-s Dielectric, piezoelectric, ferroelectric, and antiferroelectric materials

% 75. Magnetic properties and materials
% 75.80.+q Magnetomechanical and magnetoelectric effects,
% magnetostriction
% 75.85.+t Magnetoelectric effects, multiferroics

% 71.  Electronic structure of bulk materials
% 71.15.Mb Density functional theory, local density approximation,
% gradient and other corrections 

% 61. Structure of solids and liquids; crystallography
% 61.50.-f     Structure of bulk crystals
% 61.50.Ah   Theory of crystal structure, crystal symmetry; calculations and modeling 

\maketitle

\section{Introduction}

Perovskite oxide BiFeO$_3$ (BFO) continues to reveal itself as one of
the most intriguing materials of the day.  Not only does it remain the
most promising magnetoelectric multiferroic for applications at room
temperature, but it also has been shown recently to display a variety
of novel fundamental effects.\cite{catalan09} Such findings range from
an increased conductivity at specific ferroelectric domain
walls\cite{seidel09} to new structural phases in thin
films\cite{bea09,infante10} with potentially useful response
properties.\cite{zeches09,wojdel10}

The present work originated from our on-going research on enhancing the
properties of BFO by forming solid solutions such as
BiFe$_{1-x}$Co$_{x}$O$_3$\cite{azuma08} and Bi$_{1-x}${\sl RE}$_{x}$FeO$_3$
with {\sl RE}~=~La, Sm, Dy.\cite{kan10} While investigating the chemically
induced structural transitions, it became clear we needed to have a thorough
and unbiased strategy to search for possible structural phases beyond those
reported in the literature. Interestingly, when we applied such a scheme to
BFO itself, we found plenty of low-symmetry phases that are local minima of
the energy. Here we describe the lowest-energy structures that we discovered,
i.e., those most likely to be observed experimentally. We discuss the origin
of the large variety of distortions found in the calculations, and the
possibility of capturing BFO's structural richness within simple
models. Further, we comment on the implications of our findings as regards
current experimental work on BFO in both bulk and thin film forms.

%%%

\section{Methodology}

\begin{figure}
\centering
\subfigure[]{
\includegraphics[height=43mm]{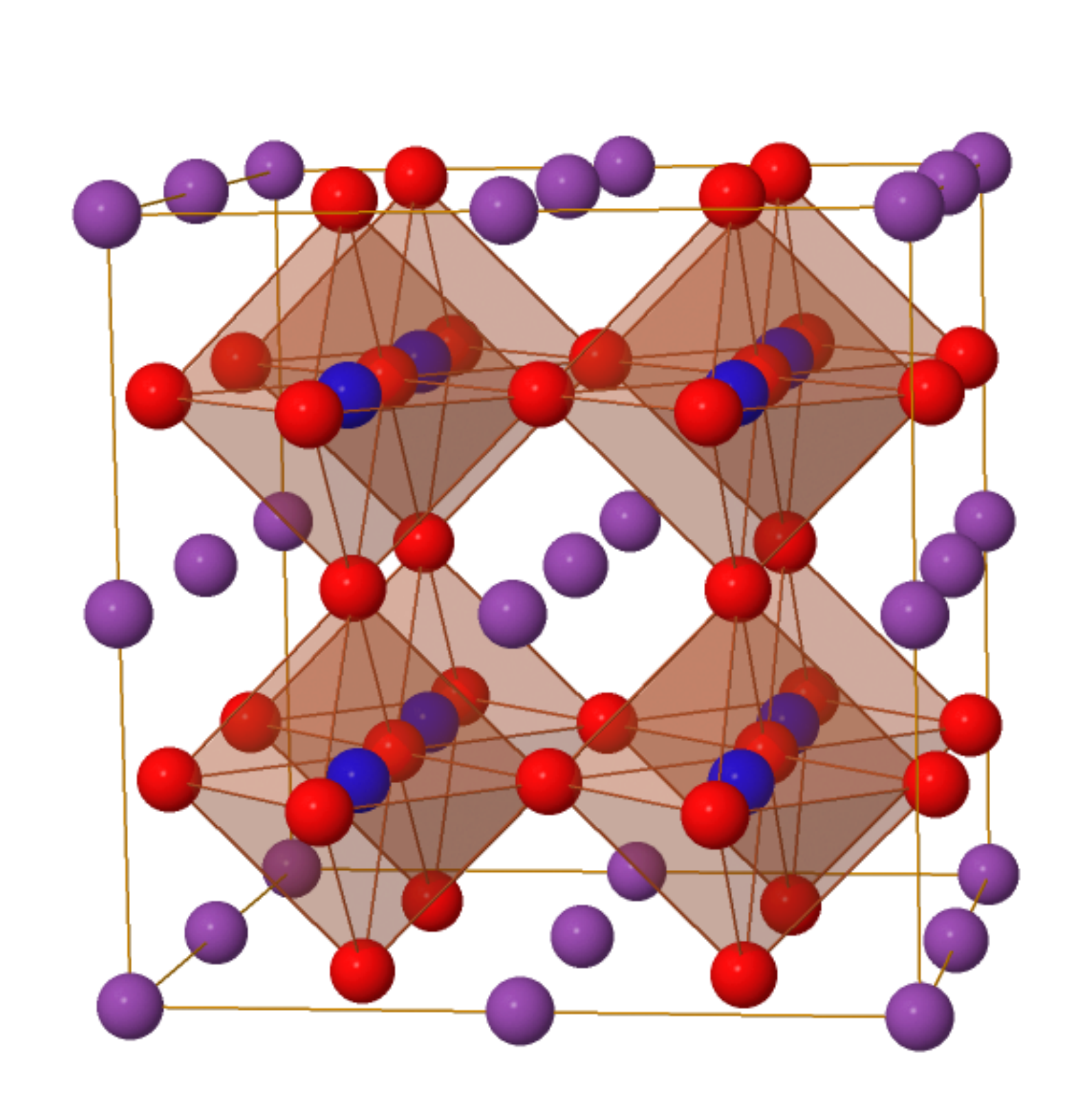}
}
\\
\subfigure[]{
\includegraphics[height=43mm]{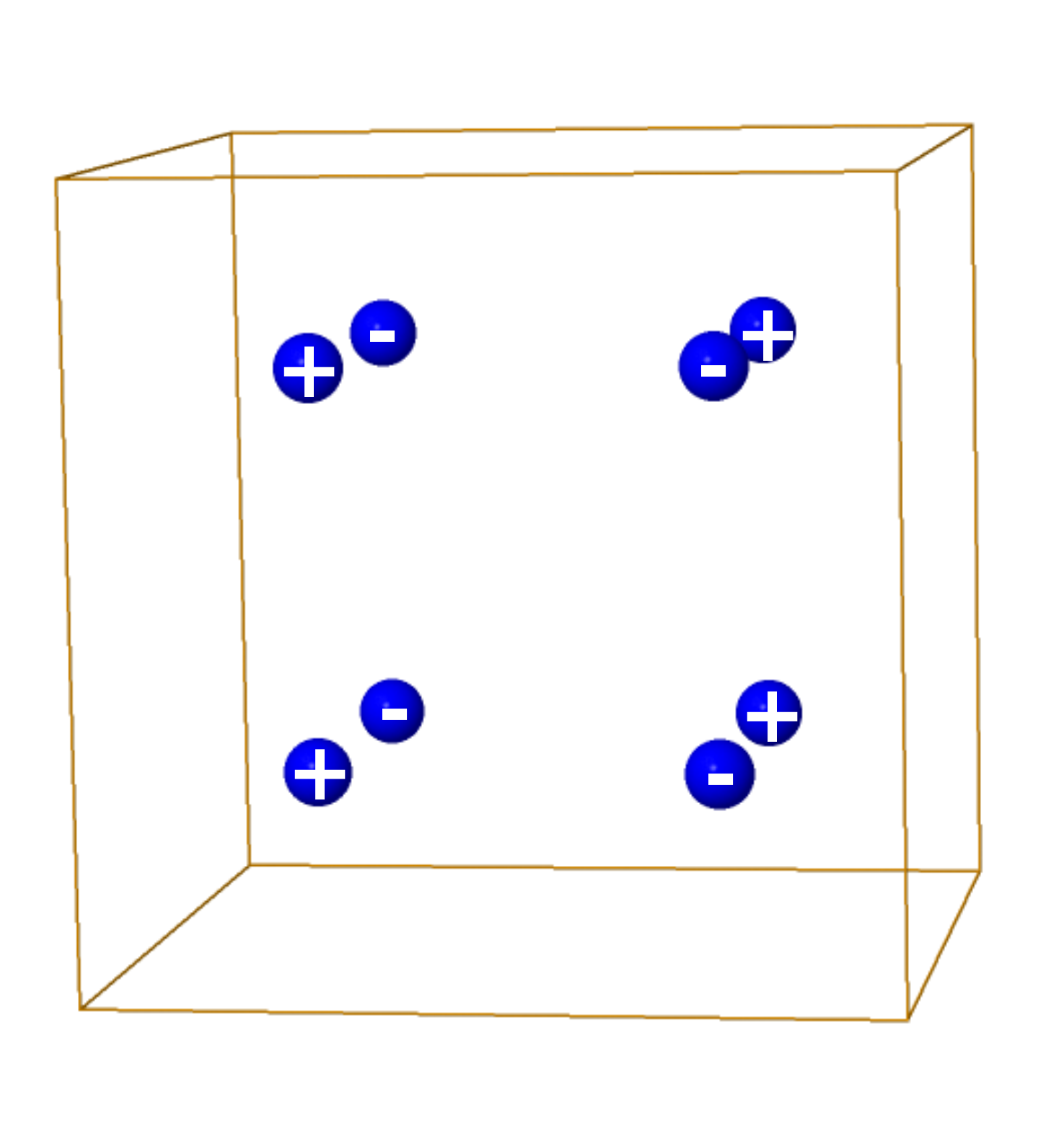}
}
\subfigure[]{
\includegraphics[height=43mm]{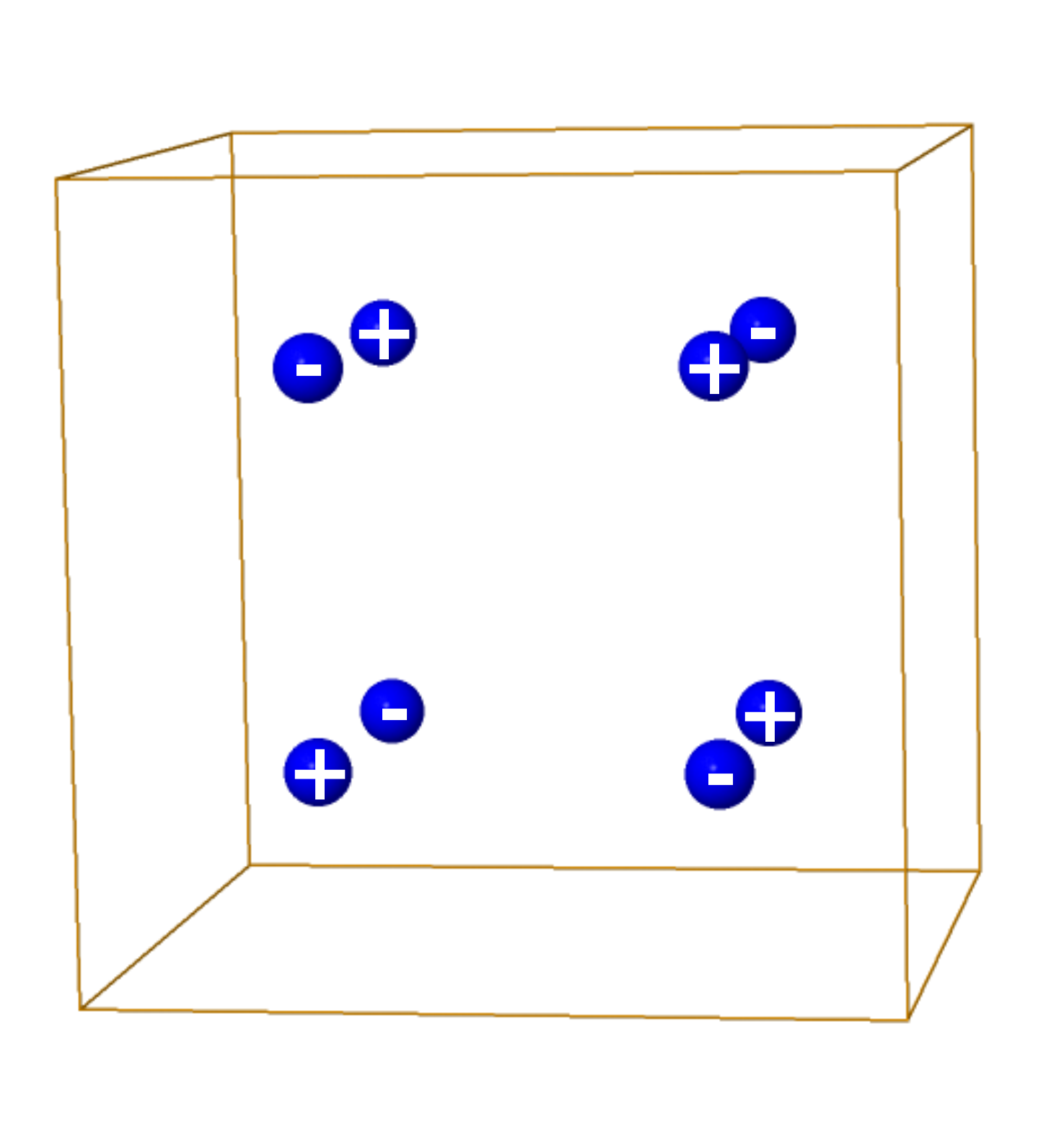}
}
\caption{(Color online.) (a) 40-atom supercell of BiFeO$_3$ (extra periodic
  images of some O and Bi atoms have been included for easier visualization).
  The O atoms occupy the vertices of the octahedra plotted, which contain Fe
  atoms at their centers; the rest of the atoms are Bi.  (b) Same cell as (a),
  illustrating a C-AFM arrangement of Fe atom spins.  (c) Same as (b), but
  with a G-AFM spin arrangement.  All the phases considered in this work have
  unit cells that are distortions of the one depicted here.}
\label{fig_1}
\end{figure}

For the simulations we used the local density (LDA\cite{lda}) and generalized
gradient (PBE\cite{perdew96} and PBEsol\cite{perdew08}) approximations to
density functional theory (DFT) as implemented in the {\sc vasp}
package.\cite{vasp} A ``Hubbard-{\sl U}'' scheme with $U=4$~eV was used for a
better treatment of iron's 3$d$ electrons;\cite{dudarev98} the corrected DFT
functionals will thus be referred to as LDA+{\sl U}, PBE+{\sl U}, and
PBEsol+{\sl U}.  We used the ``projector augmented wave'' method to represent
the ionic cores,\cite{vasp-paw} solving for the following electrons: Fe's
3$s$, 3$p$, 3$d$, and 4$s$; Bi's 5$d$, 6$s$, and 6$p$; and O's 2$s$ and
2$p$. (We checked that qualitatively correct results can be obtained without
considering semi-core electrons.) Wave functions were represented in a
plane-wave basis truncated at 500~eV, and a 2$\times$2$\times$2 $k$-point grid
was used for integrations within the Brillouin zone (BZ) corresponding to the
40-atom cell of Fig.~\ref{fig_1}. The calculation conditions were checked to
render converged results.

We worked with the 40-atom cell depicted in Fig.~\ref{fig_1}, which is
obtained by doubling the 5-atom cell of the ideal perovskite structure
along the three Cartesian directions, denoted by $x$, $y$, and $z$ in
the following. This cell is compatible with the structural distortions
that characterize the low-symmetry phases of many perovskite
oxides:~\cite{glazer72} (1) ferroelectric (FE) patterns associated
with irreducible representation $\Gamma_{4}^{-}$ (symmetry labels
correspond to the BZ of the 5-atom cubic cell); (2) anti-ferroelectric
(AFE) modes associated with zone-boundary $q$ points ($X$-like,
$M$-like, and $R$); and (3) anti-ferrodistortive (AFD) patterns
corresponding to any combination of in-phase ($M_{3}^{+}$) and
anti-phase ($R_{4}^{+}$) rotations of the O$_6$ octahedra around the
Cartesian axes. This cell is also compatible with the
anti-ferromagnetic (AFM) spin arrangements known to be most relevant
for BFO, i.e., the C-AFM and G-AFM orders sketched in
Figs.~\ref{fig_1}(b) and \ref{fig_1}(c), respectively.

To explore all these possibilities we considered a large number of starting
configurations for our structural relaxations. Specifically, we considered:
(1) all AFD patterns consisting of either an in-phase or an anti-phase
rotation around each Cartesian axis (i.e., those expressible in Glazer's
notation\cite{glazer72}); (2) various FE and AFE patterns constructed by
off-centering the Bi cations; (3) cells with cubic, tetragonal, and
orthorhombic shapes; (4) G- and C-AFM orders as well as a few attempts with
other spin arrangements. This added up to more than 300 starting
configurations. In all cases, we first ran a short molecular dynamics
simulation with random initial velocities (thus breaking all symmetries), and
then performed a full structural relaxation. We used the PBE+{\sl U}
functional for this structural search. The lowest-energy configurations
obtained were confirmed to be minima by checking their stability against ionic
and cell distortions.

\begin{table*}
\setlength{\extrarowheight}{1mm}

\caption{Energies and distortions of the most stable energy minima found, as
  well as a few saddle points (six bottom phases) included for
  reference. Columns 2-4: Energies obtained with different DFT
  functionals. Note $Pna2_1$-G goes to $Pnma$-G when relaxed with PBEsol+{\sl
    U} and LDA+{\sl U}. Columns 5-8: Distortions from the ideal cubic
  perovskite structure ($Pm\bar{3}m$) that characterize the phases. In all
  cases the FE and AFD modes fully determine the symmetry breaking. A generic
  [$x,y,z$] FE (AFD) distortion involves displacements (O$_6$ rotations) along
  (around) the $x$, $y$, and $z$ Cartesian axes. We indicate the dominant FE
  and AFD distortions in bold. Column~8 includes other modes with a
  significant contribution (at least 10\% of largest one). The mode analysis
  was done with the {\sc isodisplace} software;\protect\cite{isodisplace} note
  that $q$-points indicated in symmetry labels constitute default choices and
  do not always correspond to the actual distortion modulation (e.g., the
  $X_{5}^{+}$ and $X_{5}^{-}$ AFE modes in the Table are actually modulated
  along the $z$ direction).}
\vskip 1mm
\begin{tabular*}{2.0\columnwidth}{@{\extracolsep{\fill}}cccccccc}
\hline\hline
 & \multicolumn{3}{c}{$\Delta E = E-E(R3c$-G$)$ (meV/f.u.)} &
\multicolumn{4}{c}{Structural  distortions}\\  
\cline{2-4}
\cline{5-8}
Phase & PBE+{\sl U} & PBEsol+{\sl U} & LDA+{\sl U} & $\Gamma_{4}^{-}$ (FE) &
$R_{4}^{+}$ (AFD) & $M_{3}^{+}$ (AFD) & Additional distortions \\  
\hline
$Pc$-C   &    19  &  106   &  134
& $[x,x,\mathbf{z}]$ & $-$ & $[0,0,z]$ & AFE ($M_{5}^{-}$),
O$_6$-dist. ($\Gamma_{5}^{-}$), $c/a$=1.27
\\ 
$Cm$-C  &     12   &    103   &  132 
& $[0,y,\mathbf{z}]$ & $-$ & $[0,y,0]$ & O$_6$-dist. ($\Gamma_{5}^{-}$),
$c/a$=1.27 \\ 
$Pna2_{1}$-C &                 14 &   99    &  127 
& $[0,0,\mathbf{z}]$ &  $[x,x,0]$  & $[0,0,z \approx 0]$ & AFE ($X_{5}^{+}$,
$X_{5}^{-}$, $R_{5}^{+}$), $c/a$=1.26 \\   
$Cc$-C &                          10 &    96   &  125 
& $[x,x,\mathbf{z}]$ & $[x,x,z \approx 0]$ & $-$ &
AFE ($R_{5}^{+}$), O$_6$-dist. ($\Gamma_{5}^{-}$), $c/a$=1.25\\
$Pnma$-G       &                 60    &    27   &    14 
& $-$ & $[\mathbf{x},\mathbf{x},0]$ & $[0,0,\mathbf{z}]$ & AFE ($X_{5}^{+}$,
$R_{5}^{+}$) \\  
$Pna2_{1}$-G  &                  47    &  $-$    &   $-$ 
& $[0,0,z]$ & $[\mathbf{x},\mathbf{x},0]$ & $[0,0,\mathbf{z}]$ & AFE
($X_{5}^{+}$, $X_{5}^{-}$) \\   
$R3c$-G &                          0  &     0 &     0   
& $[\mathbf{x},\mathbf{x},\mathbf{x}]$ & $[\mathbf{x},\mathbf{x},\mathbf{x}]$
& $-$ & $-$ \\ 
\hline
$P4mm$-C        &          82   &  140   &  152 
& $[0,0,\mathbf{z}]$ & $-$ & $-$ & $c/a$=1.28 \\
$R3m$-G          &      136    &   169    &  191
& $[\mathbf{x},\mathbf{x},\mathbf{x}]$ & $-$ & $-$ & $-$ \\
$Amm2$-G       &    175    &  203 & 213
& $[\mathbf{x},\mathbf{x},0]$ & $-$ & $-$ & $-$ \\
$R\bar{3}c$-G    &   272   &   230   & 209 
& $-$  & $[\mathbf{x},\mathbf{x},\mathbf{x}]$ & $-$ & $-$ \\
$I4/mcm$-G &               430   &   372    &   344 
& $-$ & $[0,0,\mathbf{z}]$ & $-$ & $-$ \\
$Pm\bar{3}m$-G &          981  &  906  &  870     
& $-$ & $-$ & $-$ & $-$  \\
\hline\hline
\end{tabular*}
\end{table*}

\section{Results}

\subsection{Lowest-energy phases found}

Our search led to a wealth of local minima with energies in a range up
to 200~meV/f.u. above BFO's ground state. Table~I lists the
lowest-lying solutions; we show their PBE+{\sl U} energy directly
obtained from our structure search, as well as the energies obtained
by relaxing the PBE+{\sl U} structure using the PBEsol+{\sl U} and
LDA+{\sl U} functionals. Note that the energy differences between
phases are strongly dependent on the DFT functional; we will address
this issue below.  Table~I also includes a short description of the
phases found, which we label by their atomic space group and type of
AFM order (e.g., $R3c$-G for the ground state); the complete
structural information and computed polarization
values\cite{fn:polarization} are given in Tables II and III. Let us
note that our work with BFO and other compounds confirms that PBEsol
is more accurate than PBE and LDA for predicting the atomic structure
of individual phases.\cite{perdew08} Thus, the crystallographic data
reported here correspond to PBEsol+{\sl U}-relaxed
structures. Finally, Fig.~2 shows sketches of the structures obtained,
and the most relevant distortion modes are depicted in Fig.~3.

\begin{figure*}
\centering
\subfigure[~$Pc$-C]{
\includegraphics[width=35mm]{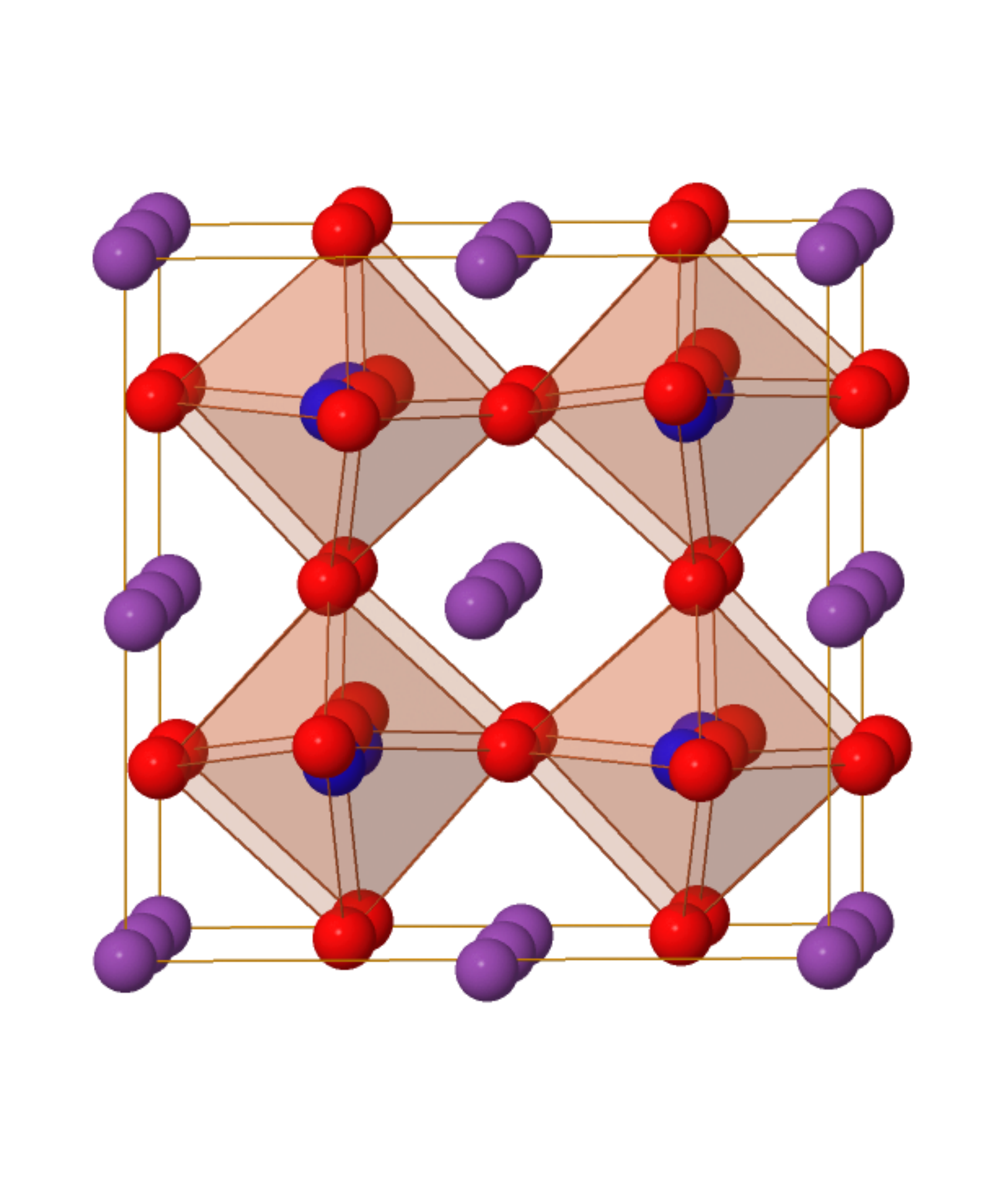}
\includegraphics[width=35mm]{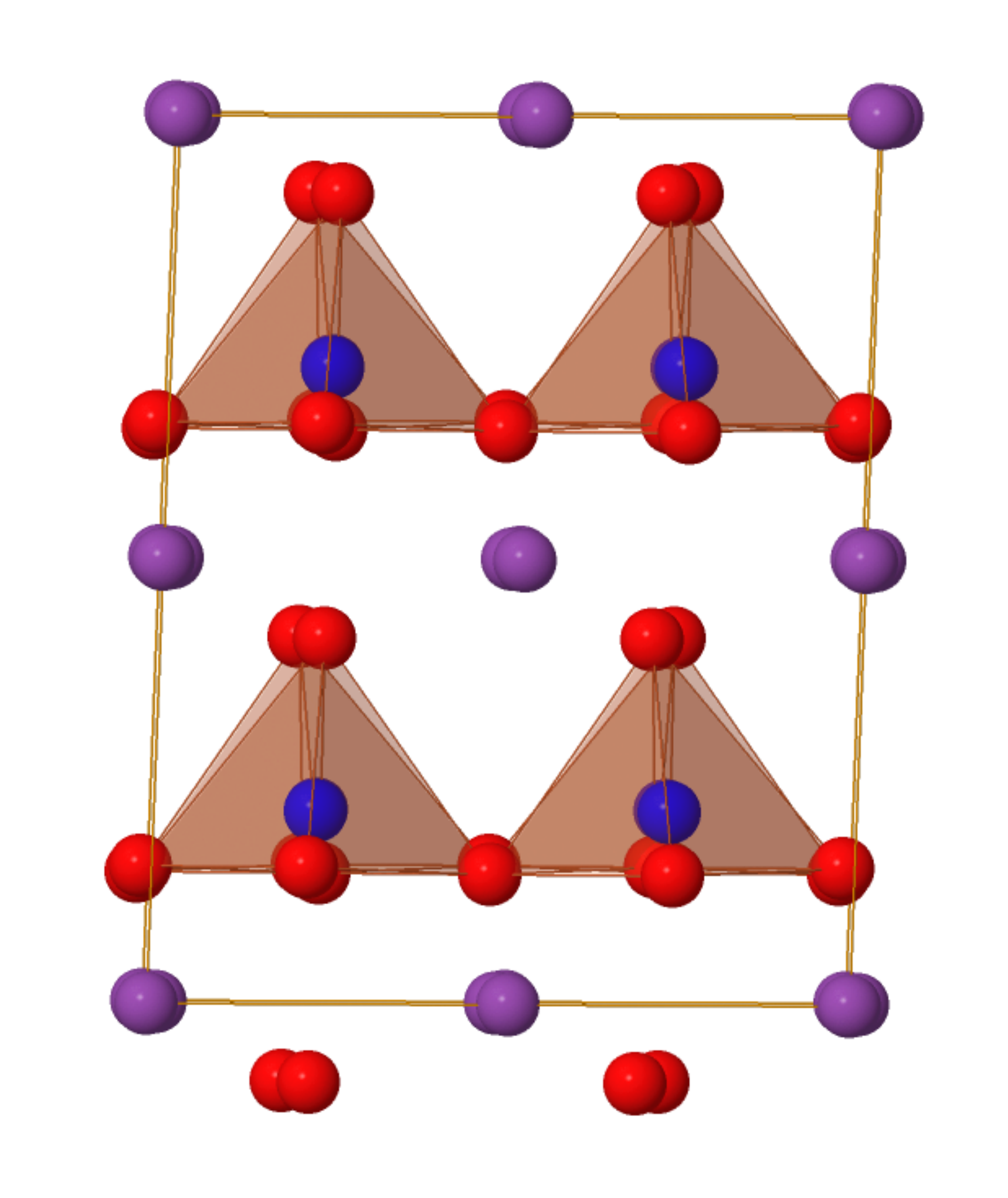}
}
\hspace{1cm}
\subfigure[~$Cm$-C]{
\includegraphics[width=35mm]{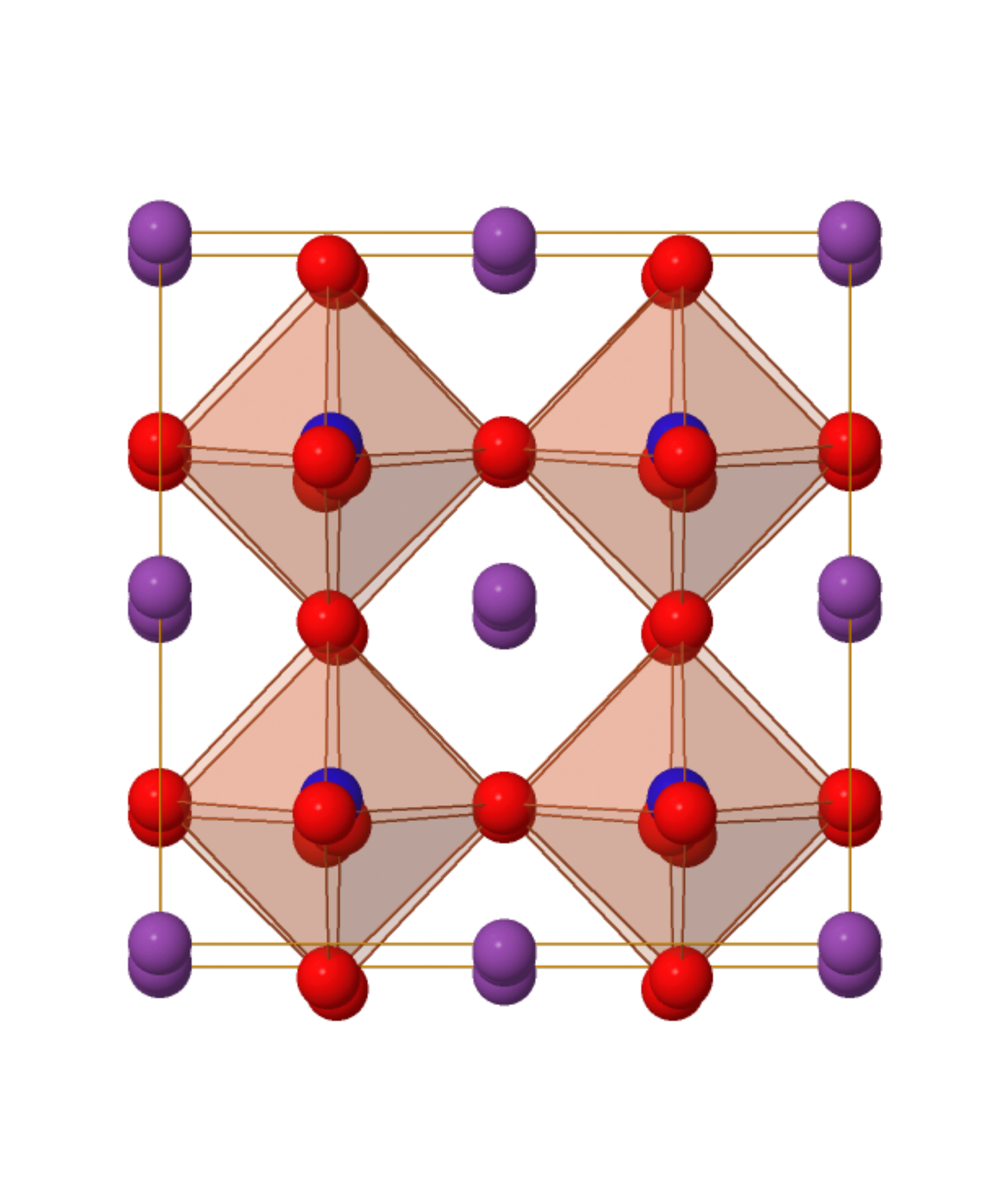}
\includegraphics[width=35mm]{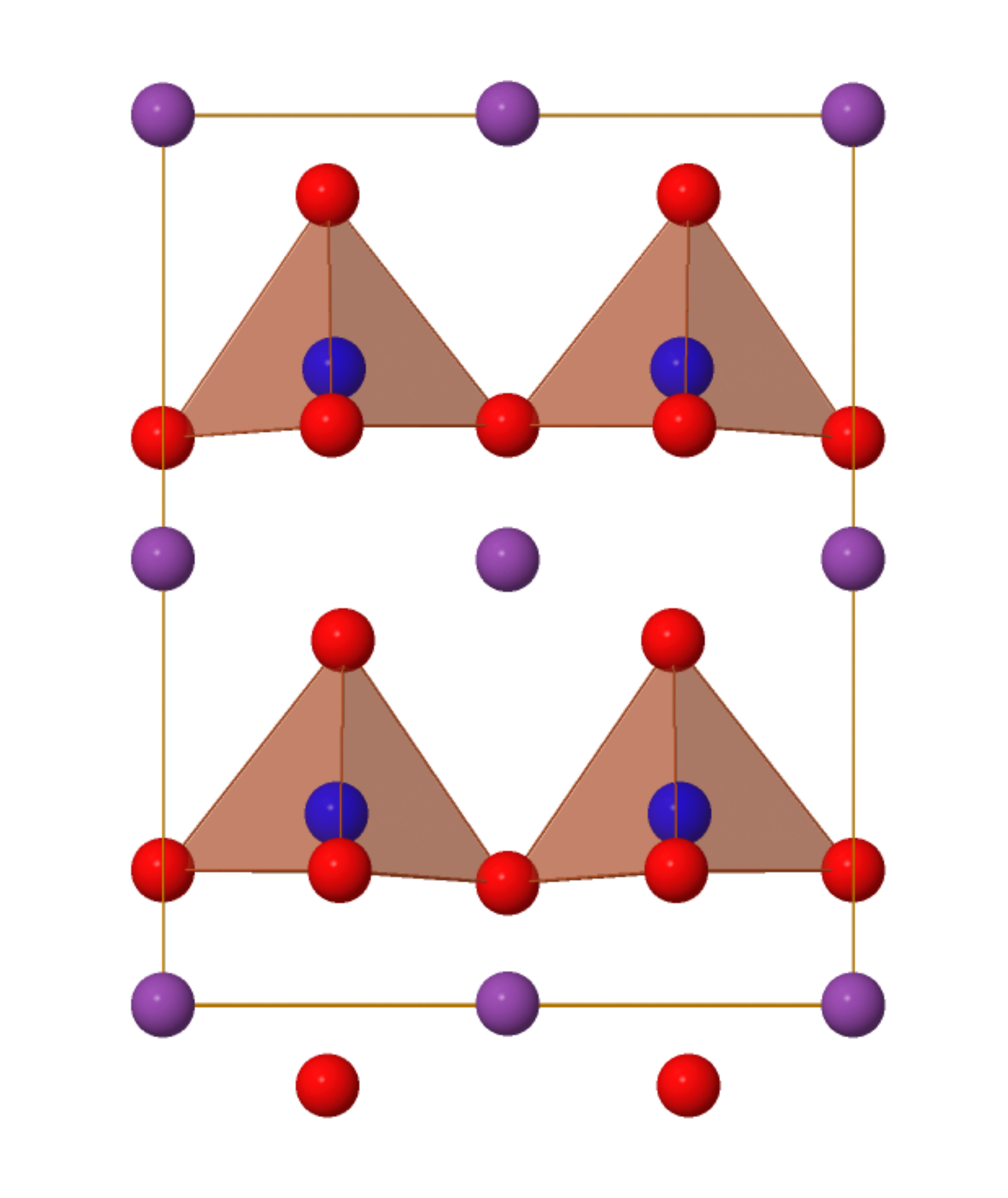}
}
\\
\subfigure[~$Pna2_1$-C]{
\includegraphics[width=35mm]{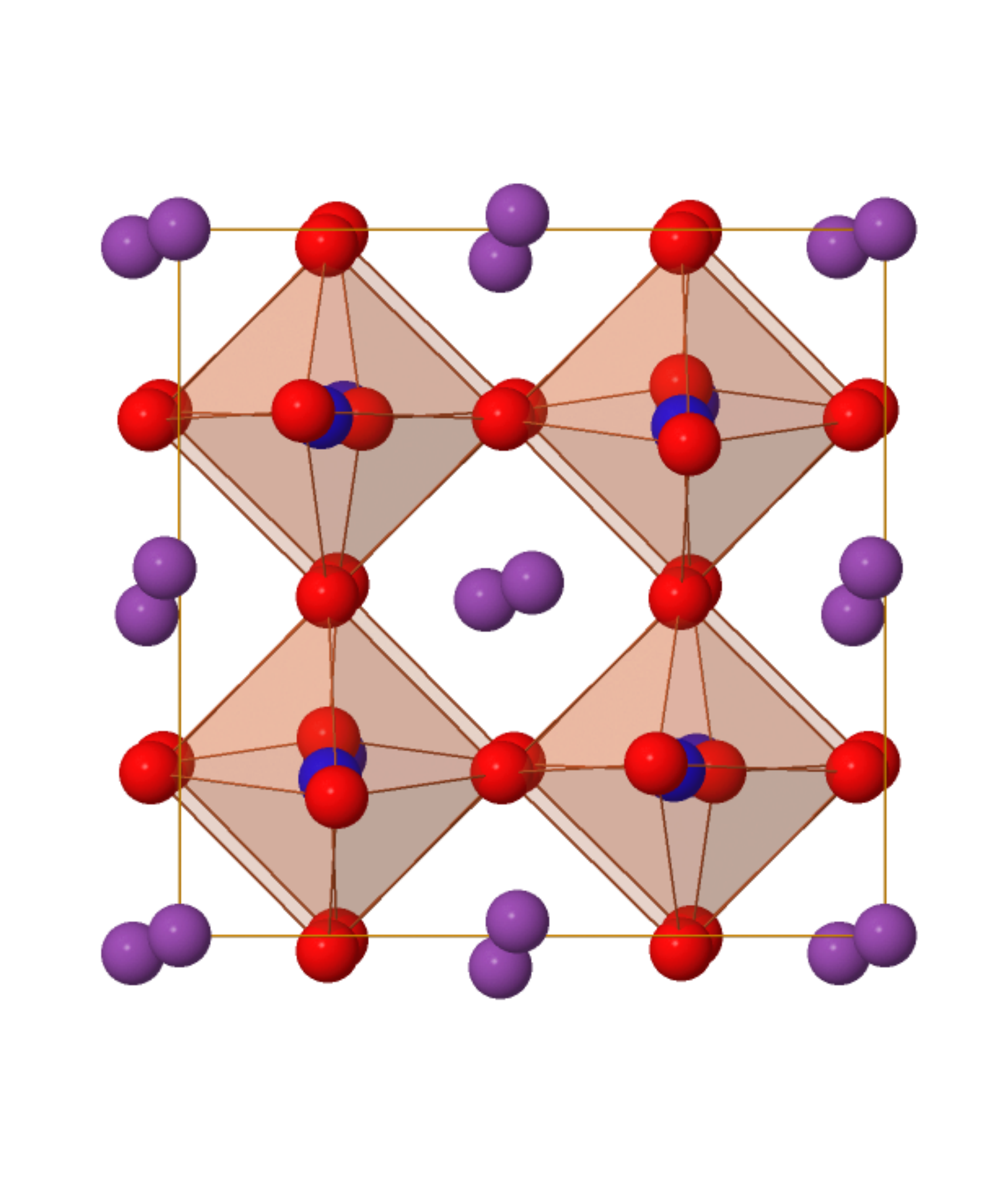}
\includegraphics[width=35mm]{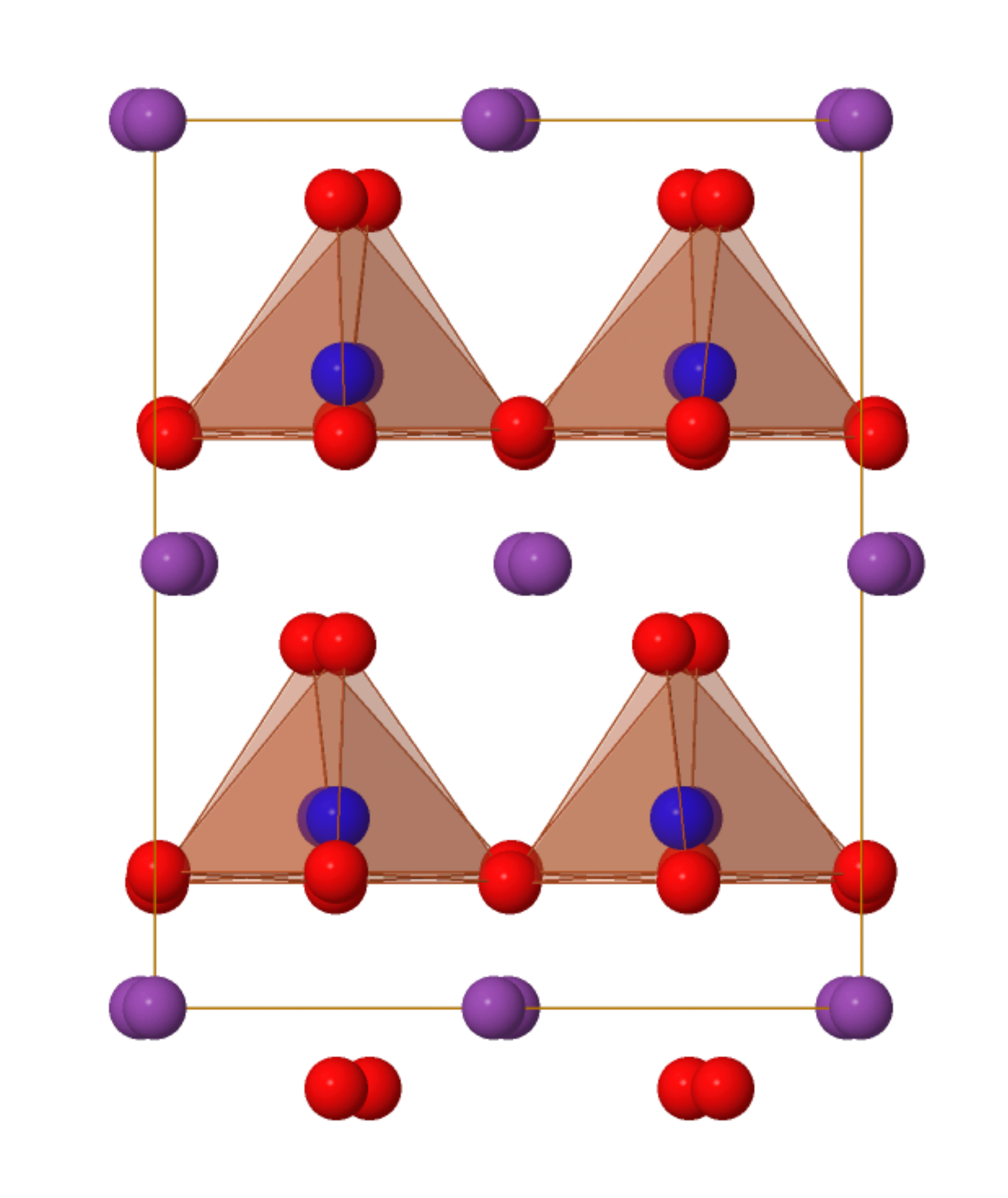}
}
\hspace{1cm}
\subfigure[~$Cc$-C]{
\includegraphics[width=35mm]{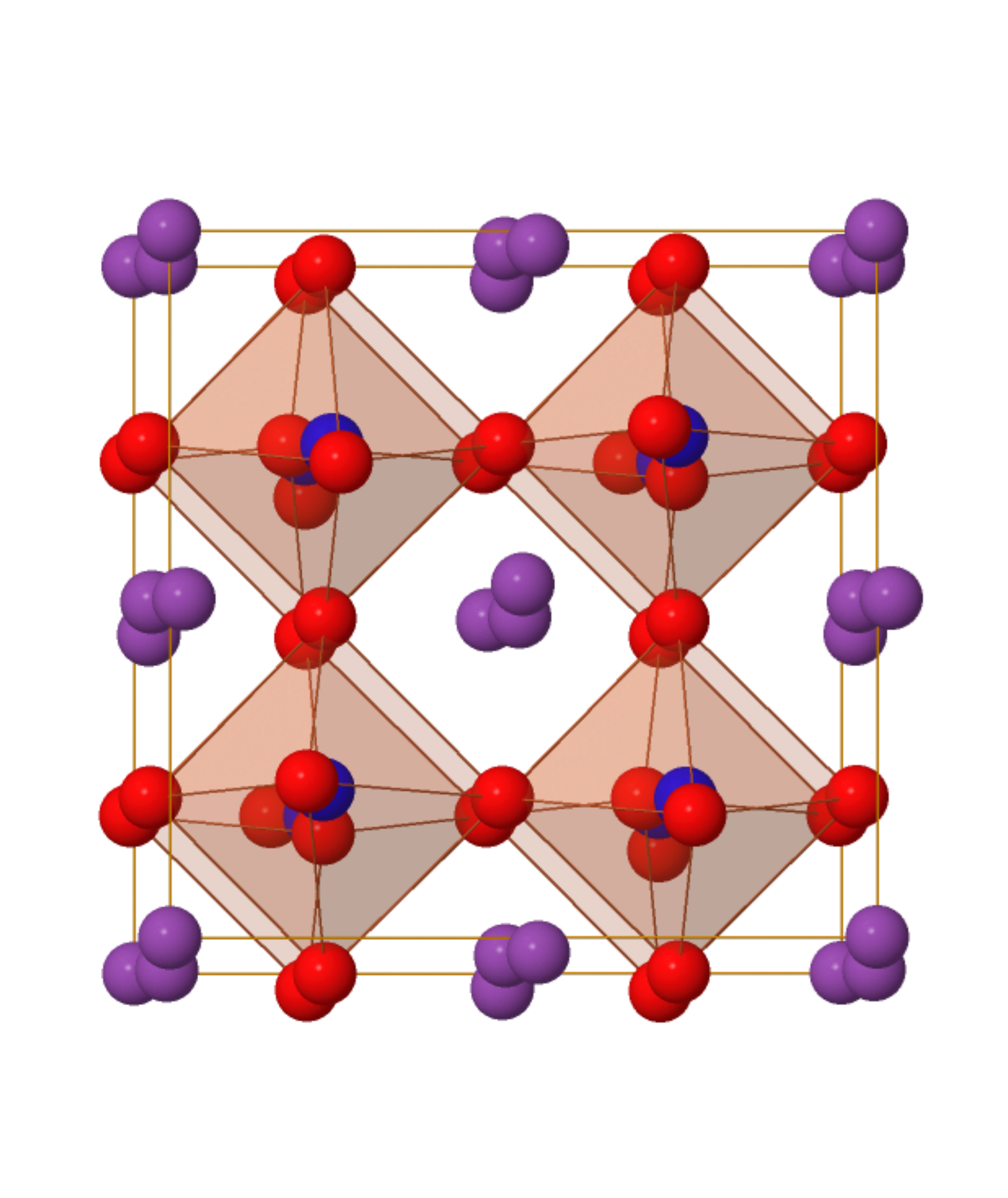}
\includegraphics[width=35mm]{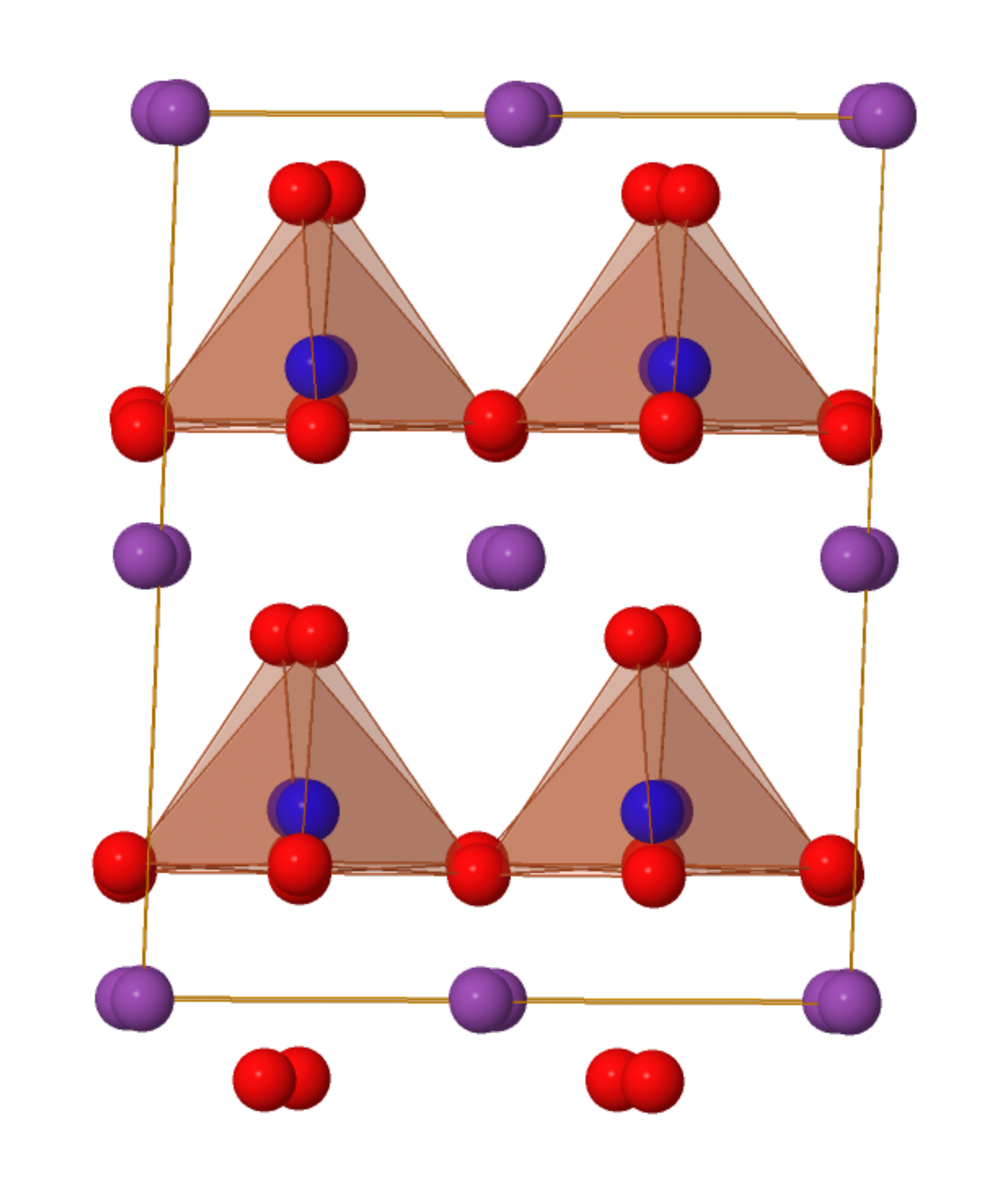}
}
\\
\subfigure[~$Pnma$-G]{
\includegraphics[width=35mm]{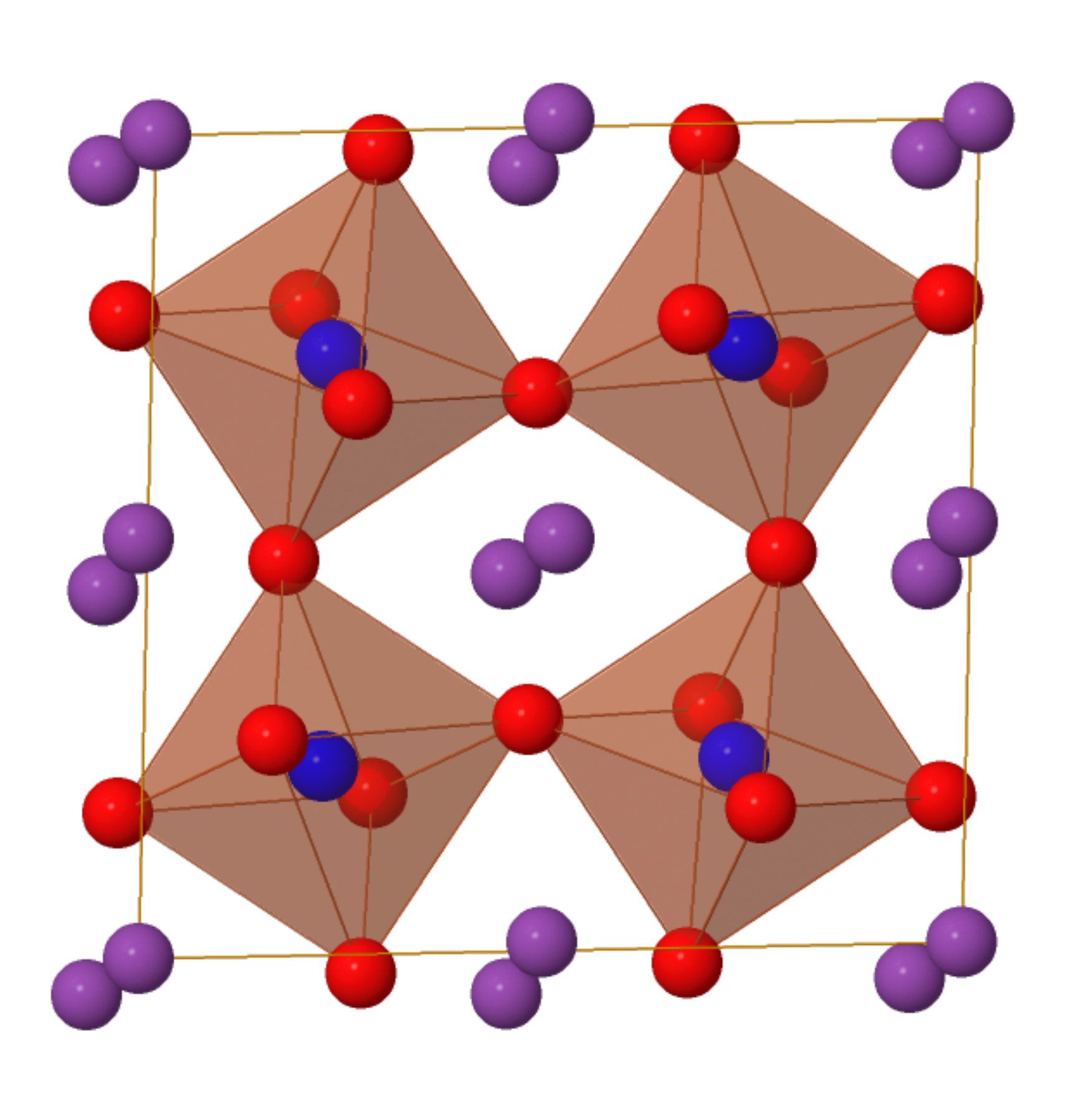}
\includegraphics[width=35mm]{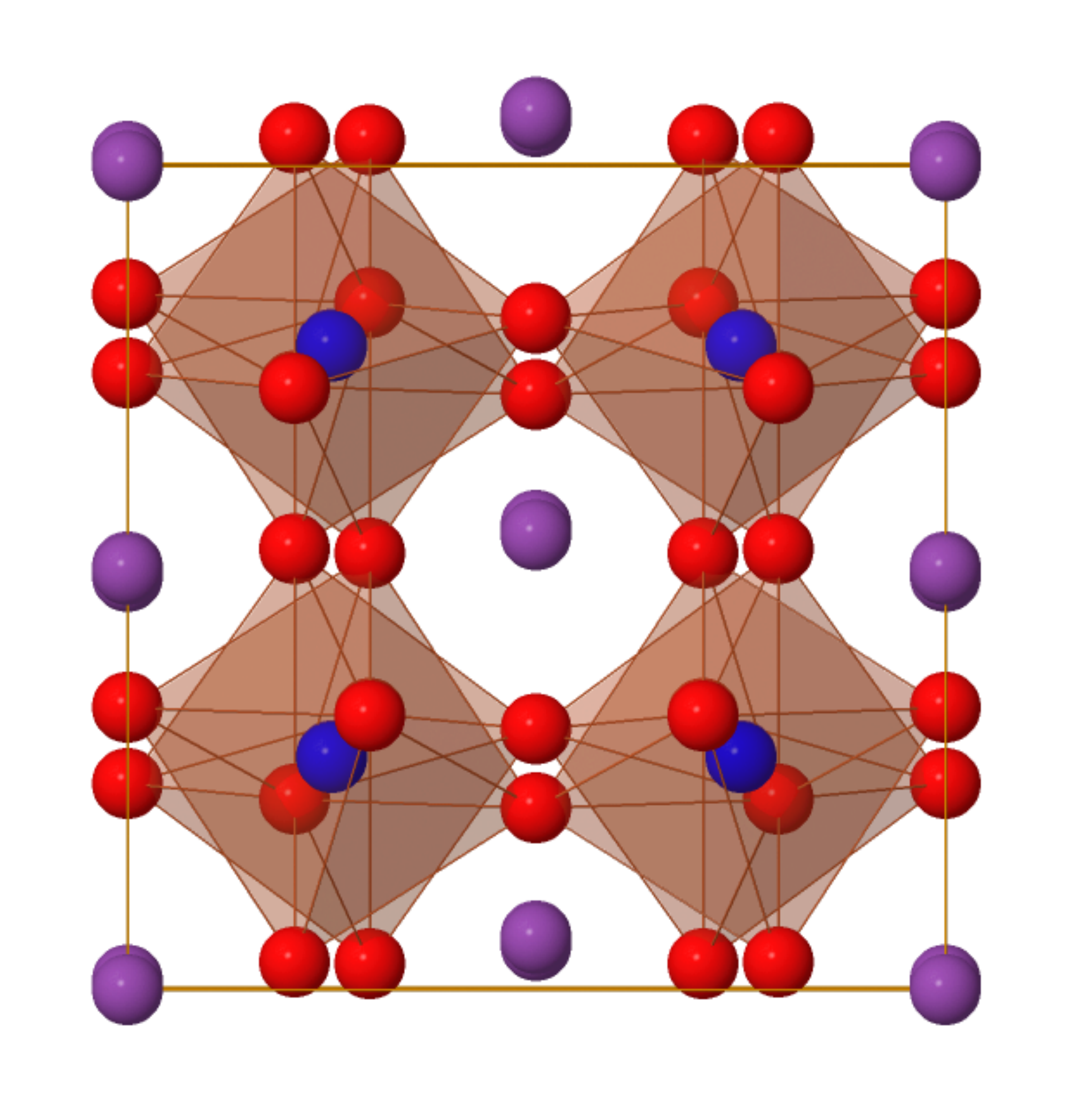}
}
\hspace{1cm}
\subfigure[~$R3c$-G]{
\includegraphics[width=35mm]{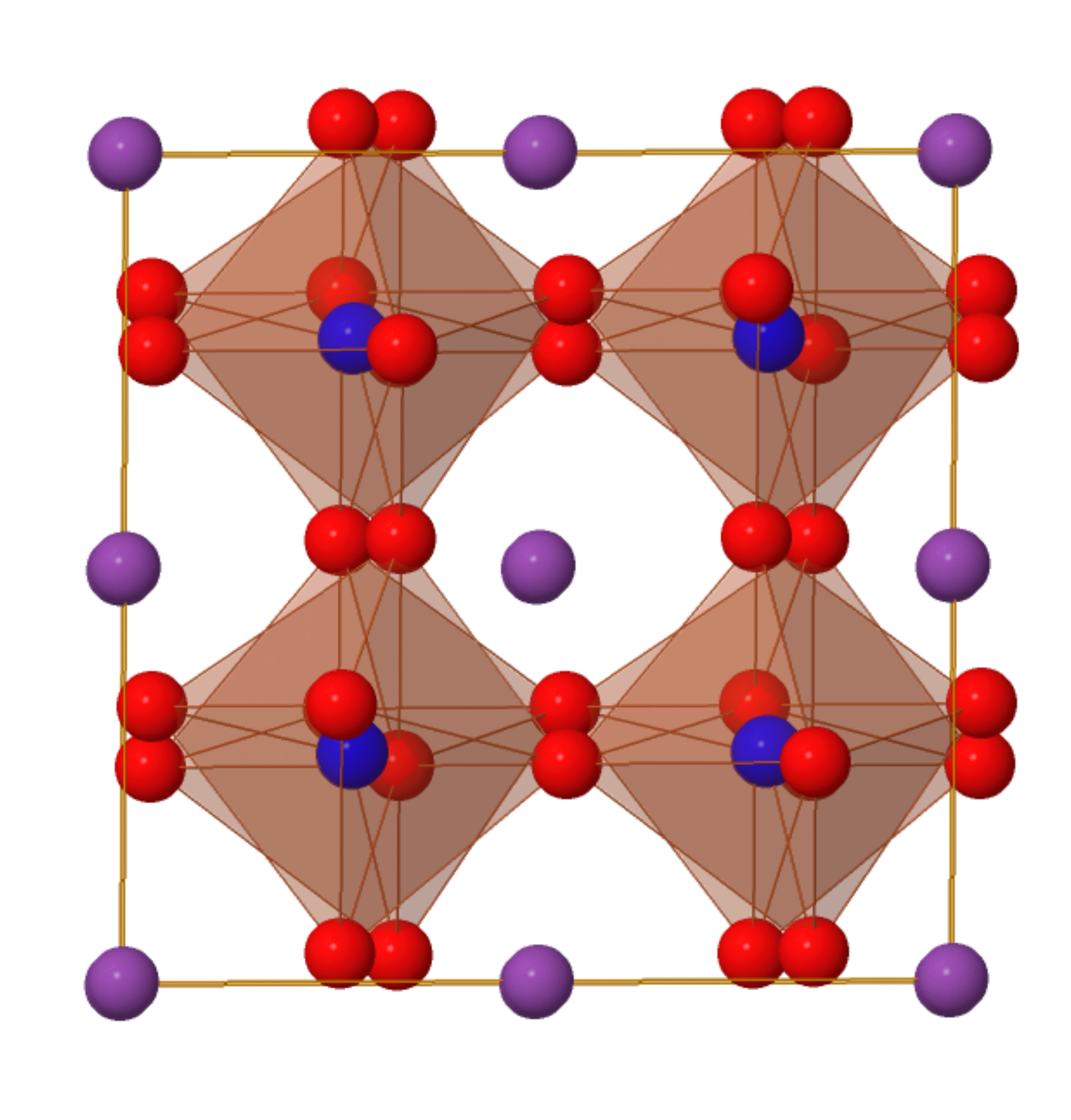}
}
\caption{(Color online.) Energy minimum configurations obtained.
  (a)-(d) C-AFM super-tetragonal phases; in the left (right) image the
  $c$ axis is perpendicular (parallel) to the page.  (e)-(f) G-AFM
  phases; two pseudo-cubic axis are equivalent in (e), with the left
  (right) figure having the non-equivalent axis perpendicular
  (parallel) to the page; the three pseudo-cubic axis are equivalent in
  (f).  The atomic species can be identified as in
  Fig.~\protect\ref{fig_1}.}
\label{fig_2}
\end{figure*}

\begin{figure}
\centering
\subfigure[~$\Gamma_4^-$]{
\includegraphics[width=27mm]{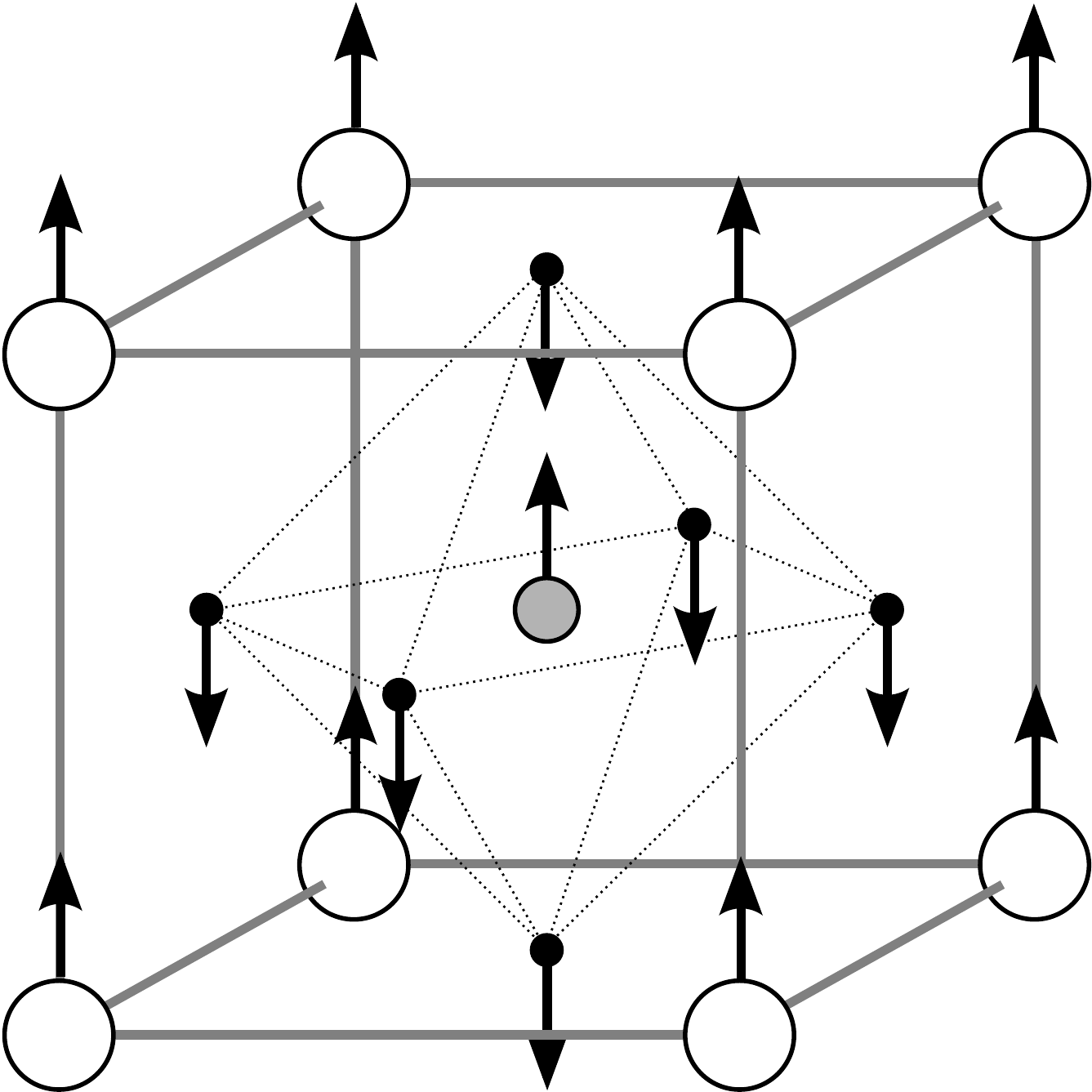}
}
\hspace{1cm}
\subfigure[~$\Gamma_5^-$]{
\includegraphics[width=27mm]{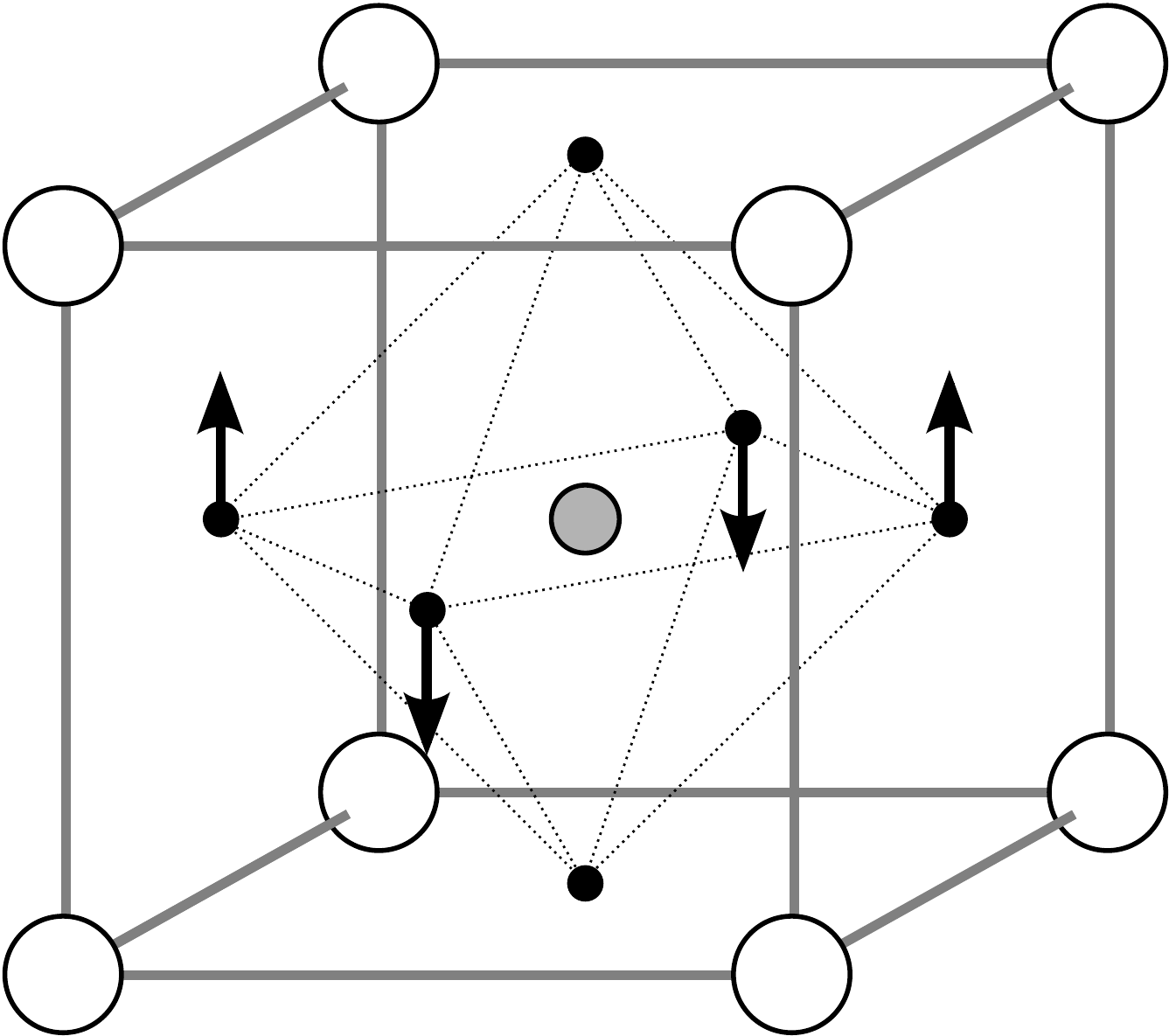}
}
\subfigure[~${\rm X}_5^+$]{
\includegraphics[width=29mm]{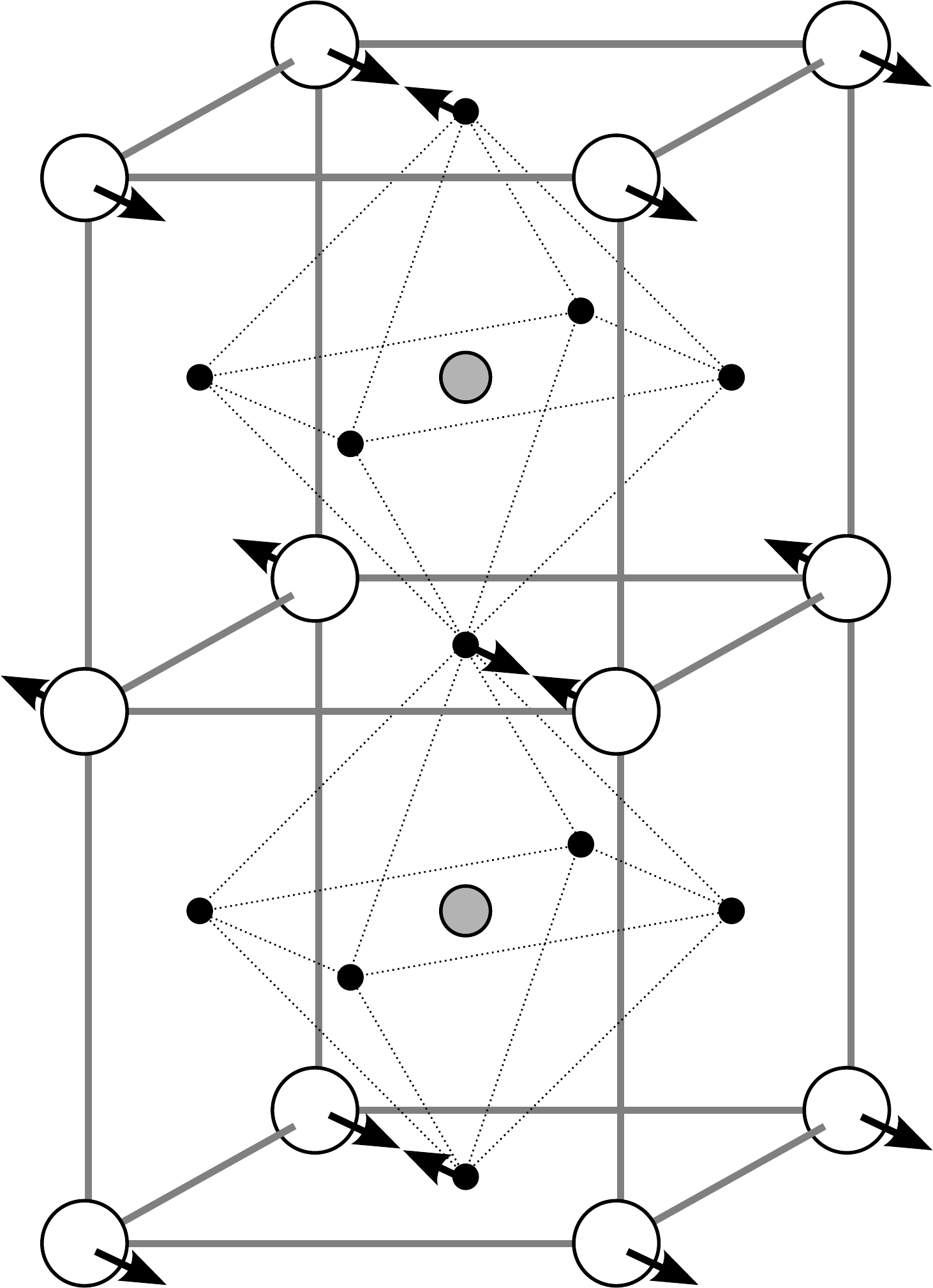}
}
\hspace{1cm}
\subfigure[~${\rm X}_5^-$]{
\includegraphics[width=27mm]{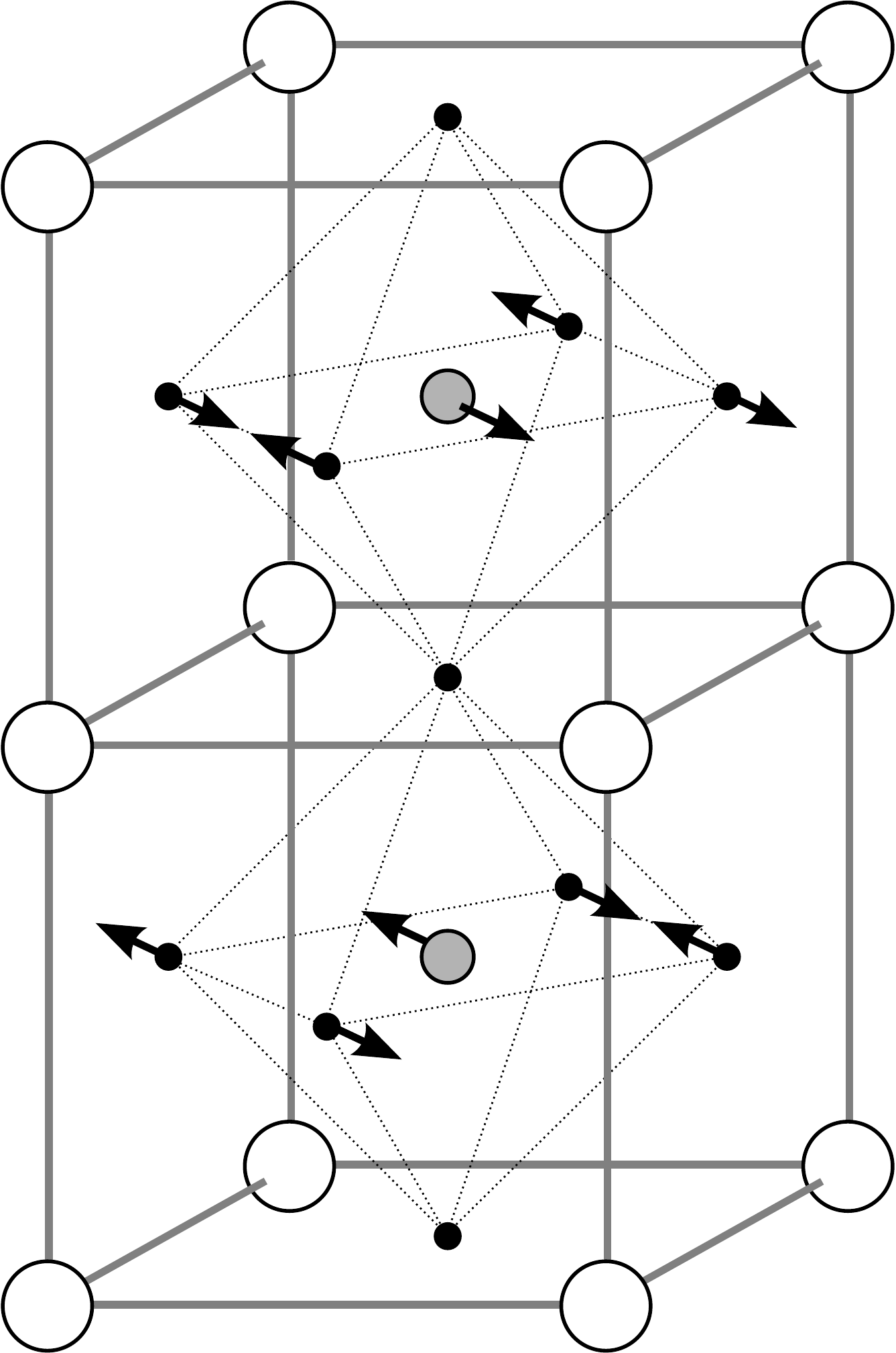}
}
\subfigure[~${\rm M}_5^-$]{
\includegraphics[width=38mm]{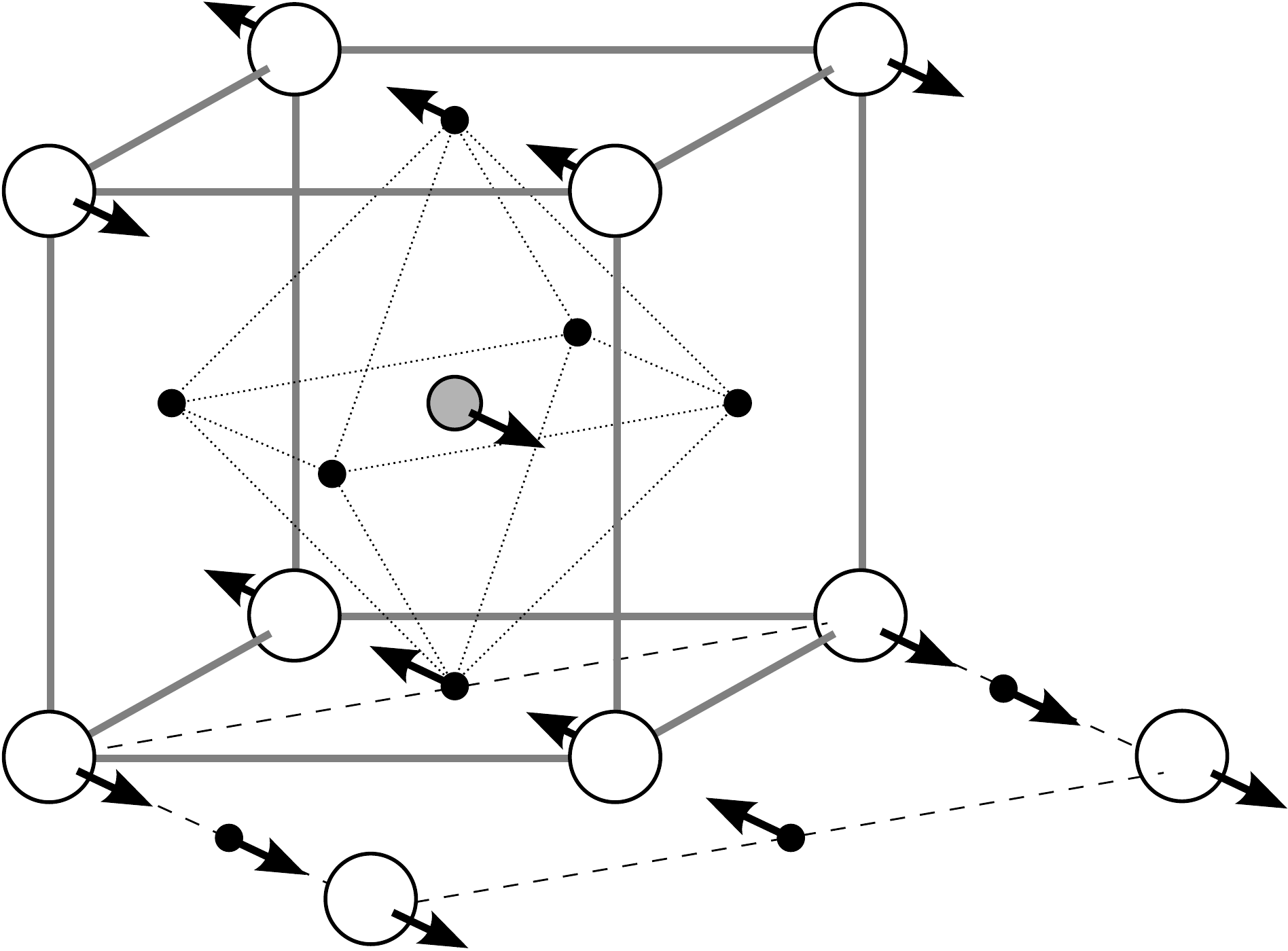}
}
\subfigure[~${\rm R}_5^+$]{
\includegraphics[width=38mm]{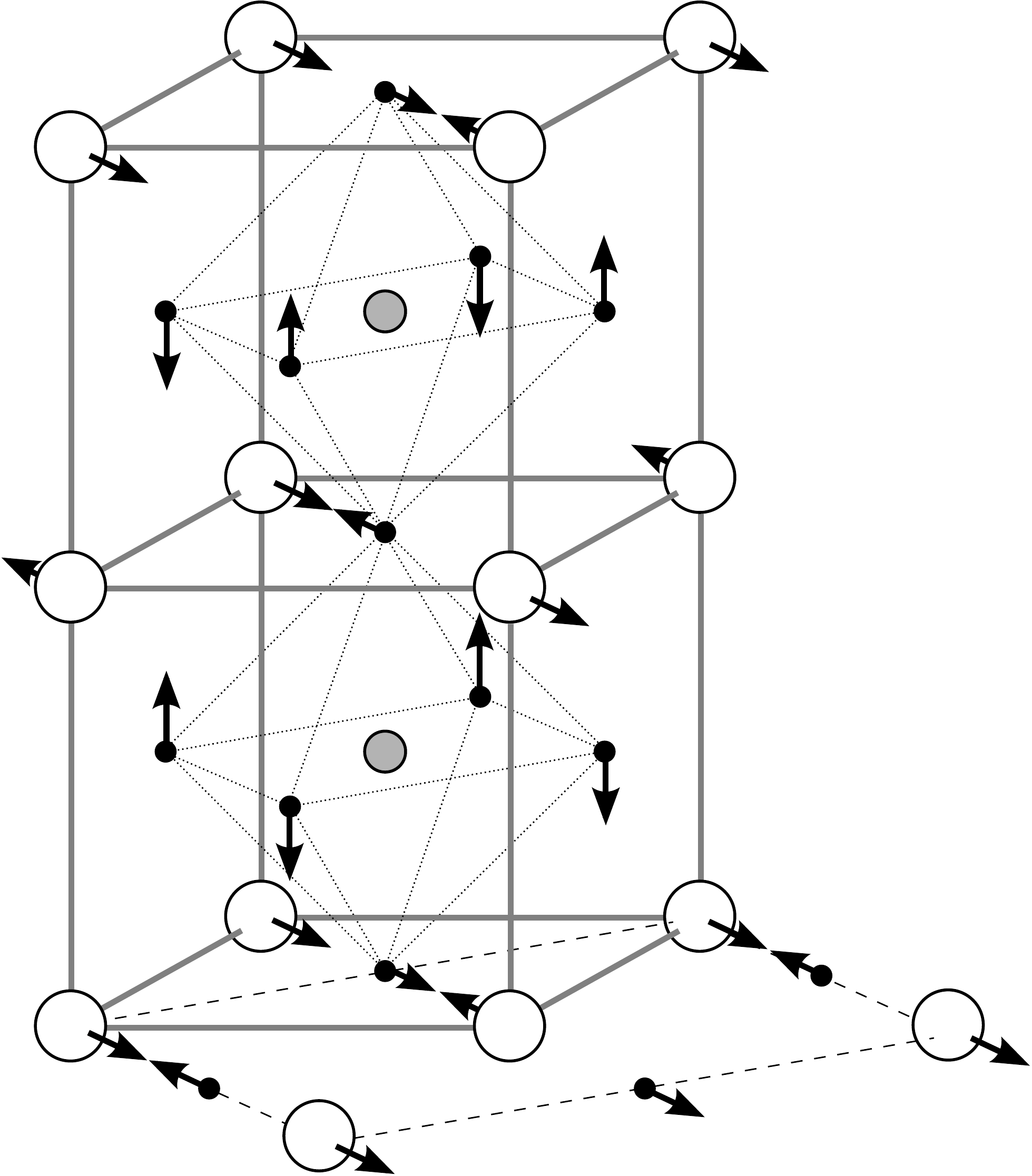}
}
\caption{Illustration of atomic displacements for different symmetry
  modes of BFO: (a) soft FE mode, (b)-(f) secondary modes mentioned in
  Table I.  Only displacement directions, not magnitudes, are
  indicated; for the (a) case, the PBEsol+{\sl U} computed atomic
  displacements are quoted in the caption of Fig.~\ref{fig_7}.
  White, grey, and black circles represent Bi, Fe, and O atoms,
  respectively.  }
\label{fig_3}
\end{figure}

All the functionals correctly predict the $R3c$ phase with G-AFM spin order as
the ground state of BFO. This phase displays a spontaneous polarization along
the [111] Cartesian direction, and anti-phase O$_6$ rotations around the same
axis ($a^{-}a^{-}a^{-}$ in Glazer's notation).

We also found two orthorhombic phases that are
similar to $R3c$-G in that they involve a relatively small distortion
of the ideal cubic cell and favor the G-AFM order: $Pnma$-G and
$Pna2_{1}$-G. 

The $Pnma$-G structure is paraelectric (PE). As shown in Table~I, it is
characterized by an O$_6$ rotation pattern ($a^{-}a^{-}b^{+}$) that involves
anti-phase rotations around [110] and in-phase rotations around [001]. This
phase is the ground state of many perovskites, LaFeO$_3$ being the most
relevant one for the current discussion. Interestingly, BFO's $Pnma$-G phase
can be aptly described as AFE, because the Bi cations present large anti-polar
displacements in the (001) plane (associated with the $X_{5}^{+}$ mode of
Table~I and Fig.~3); the computed off-centering of the Bi cations is about
0.3~\AA. (Such an AFE pattern is allowed by symmetry in LaFeO$_3$ too; in that
case we obtain La off-centers by about 0.2~\AA.)

The $Pna2_{1}$-G phase is similar to $Pnma$-G, but with an additional FE
distortion along the axis of the in-phase rotations. As compared with that of
$Pnma$-G, the 40-atom cell of the $Pna2_{1}$-G phase is elongated along the
polarization direction; this reflects the usual coupling between the FE
distortion and strain observed in perovskite oxides.

Regarding magnetism, the $R3c$-G, $Pnma$-G, and $Pna2_{1}$-G phases
display strong AFM exchange couplings between neighboring Fe ions, as
evidenced by a large energy splitting of more than
200~meV/f.u. between the G-AFM and ferromagnetic (FM)
configurations. This is consistent with the high magnetic ordering
temperature observed in bulk BFO.

\begin{table*}
\setlength{\extrarowheight}{1mm}

\caption{Computed PBEsol+{\sl U} lattice parameters (corresponding to the
  40-atom cell of Fig.~1) and polarization values for the six stable
  phases of BFO listed in Table~I. The polarization direction is given in
  a Cartesian reference that corresponds almost exactly with the 40-atom
  cell vectors. For comparison, we also include the result for the $P4mm$-C
  structure.}
\vskip 1mm
\begin{tabular*}{2.0\columnwidth}{@{\extracolsep{\fill}}ccccccccc}
\hline\hline
 & \multicolumn{6}{c}{Lattice parameters} &
\multicolumn{2}{c}{Polarization}\\  
\cline{2-7}
\cline{8-9}
Phase & $a$ (\AA) & $b$ (\AA) & $c$ (\AA)
      & $\alpha$ ($^\circ$) & $\beta$ ($^\circ$) & $\gamma$ ($^\circ$) 
      & Magnitude (C/m$^2$) & Direction \\
\hline
$Pc$-C        & 7.500 & 7.500 & 9.489  &  88.1 & 88.1 & 89.7
              & 1.20 & (0.29, 0.29, 0.92)  \\
$Cm$-C        & 7.380 & 7.608 & 9.533  &  86.6 & 90.0 & 90.0
              & 1.50 & (0.00, 0.30, 0.95)  \\
$Pna2_{1}$-C  & 7.515 & 7.515 & 9.452  &  90.0 & 90.0 & 90.0
              & 1.39 & (0.00, 0.00, 1.00)  \\
$Cc$-C        & 7.527 & 7.527 & 9.444  &  88.0 & 88.0 & 90.0
              & 1.45 & (0.23, 0.23, 0.94)  \\
$Pnma$-G      & 7.830 & 7.830 & 7.770  &  90.0 & 90.0 & 87.6
              &   0 & - \\
$R3c$-G       & 7.893 & 7.893 & 7.893  &  89.5 & 89.5 & 89.5
              &  0.91 & (0.58, 0.58, 0.58)  \\
\hline
$P4mm$-C      & 7.414 & 7.414 & 9.526  &  90.0 & 90.0 & 90.0
              & 1.52 & (0.00, 0.00, 1.00)  \\
\hline\hline
\end{tabular*}
\end{table*}

\begin{table}
\setlength{\extrarowheight}{1mm}

\caption{Energy minima structures of Table~I as obtained from PBEsol+{\sl
    U} calculations. In the case of the $Pna2_{1}$-G phase, the
  PBE+{\sl U} result is given (see text).}

\vskip 1mm

\begin{tabular*}{0.95\columnwidth}{@{\extracolsep{\fill}}ll}
\hline\hline
$Pc$-C          & $a$~=~7.291~\AA \;\; $b$~=~5.291~\AA \;\; $c$~=~5.315~\AA \\
(unique axis $b$) & $\alpha$~=~$\gamma$~=~90$^{\circ}$ \;\; $\beta$~=~139.46$^{\circ}$ \\  [1ex]
\hline
\end{tabular*}
\begin{tabular*}{0.95\columnwidth}{@{\extracolsep{\fill}}ccccc}
Atom & Wyc. & $x$    & $y$    & $z$    \\
Bi   & 2a   & 0.8692 & 0.2649 & 0.4158 \\
Fe   & 2a   & 0.4372 & 0.2467 & 0.4361 \\
O    & 2a   & 0.0471 & 0.7150 & 0.5161 \\
O    & 2a   & 0.5781 & 0.5084 & 0.3342 \\
O    & 2a   & 0.5609 & 0.0152 & 0.2979 \\ [1ex]
\end{tabular*}

\begin{tabular*}{0.95\columnwidth}{@{\extracolsep{\fill}}ll}
\hline\hline
$Cm$-C          & $a$~=~9.534~\AA \;\; $b$~=~7.380~\AA \;\; $c$~=~3.804~\AA \\
(unique axis $b$) & $\alpha$~=~$\gamma$~=~90$^{\circ}$ \;\; $\beta$~=~86.60$^{\circ}$ \\  [1ex]
\hline
\end{tabular*}
\begin{tabular*}{0.95\columnwidth}{@{\extracolsep{\fill}}ccccc}
Atom & Wyc. & $x$    & $y$    & $z$    \\
Bi   & 2a   & 0.4948 & 0      & 0.9617 \\
Bi   & 2a   & 0.9959 & 0      & 0.9418 \\
Fe   & 4b   & 0.2810 & 0.2482 & 0.5184 \\
O    & 2a   & 0.3590 & 0      & 0.5151 \\
O    & 2a   & 0.8446 & 0      & 0.5261 \\
O    & 4b   & 0.0864 & 0.2388 & 0.5689 \\
O    & 4b   & 0.3449 & 0.2443 & 0.0153 \\ [1ex]
\end{tabular*}

\begin{tabular*}{0.95\columnwidth}{@{\extracolsep{\fill}}ll}
\hline\hline
$Pna2_{1}$-C        & $a$~=~5.314~\AA \;\; $b$~=~5.314~\AA \;\; $c$~=~9.452~\AA \\
                & $\alpha$~=~$\beta$~=~$\gamma$~=~90$^{\circ}$ \\  [1ex]
\hline
\end{tabular*}
\begin{tabular*}{0.95\columnwidth}{@{\extracolsep{\fill}}ccccc}
Atom & Wyc. & $x$    & $y$    & $z$    \\
Bi   & 4a   & 0.5451 & 0.4799 & 0.4590 \\
Fe   & 4a   & 0.0195 & 0.5127 & 0.2448 \\
O    & 4a   & 0.0357 & 0.5476 & 0.0493 \\
O    & 4a   & 0.2669 & 0.7524 & 0.3170 \\
O    & 4a   & 0.2633 & 0.2491 & 0.3058 \\ [1ex]
\end{tabular*}

\begin{tabular*}{0.95\columnwidth}{@{\extracolsep{\fill}}ll}
\hline\hline
$Cc$-C        & $a$~=~10.604~\AA \;\; $b$~=~5.322~\AA \;\; $c$~=~5.323~\AA \\
(unique axis $b$)                & $\alpha$~=~$\gamma$~=~90$^{\circ}$ \;\; $\beta$ = 62.80$^{\circ}$ \\  [1ex]
\hline
\end{tabular*}
\begin{tabular*}{0.95\columnwidth}{@{\extracolsep{\fill}}ccccc}
Atom & Wyc. & $x$    & $y$    & $z$    \\
Bi   & 4a   & 0.4829 & 0.7707 & 0.1102 \\
Fe   & 4a   & 0.2689 & 0.2630 & 0.2799 \\
O    & 4a   & 0.0727 & 0.2986 & 0.4448 \\
O    & 4a   & 0.3290 & 0.9986 & 0.4671 \\
O    & 4a   & 0.3405 & 0.5032 & 0.4593 \\ [1ex]
\hline\hline
\end{tabular*}
\end{table}

\setcounter{table}{2}
\begin{table}
\setlength{\extrarowheight}{1mm}

\caption{(contd.)}

\vskip 1mm

\begin{tabular*}{0.95\columnwidth}{@{\extracolsep{\fill}}ll}
\hline\hline
$Pnma$-G        & $a$~=~5.650~\AA \;\; $b$~=~7.770~\AA \;\; $c$~=~5.421~\AA \\
                & $\alpha$~=~$\beta$~=~$\gamma$~=~90$^{\circ}$ \\  [1ex]
\hline
\end{tabular*}
\begin{tabular*}{0.95\columnwidth}{@{\extracolsep{\fill}}ccccc}
Atom & Wyc. & $x$    & $y$    & $z$    \\
Bi   & 4c   & 0.0523 & 1/4    & 0.0100 \\
Fe   & 4b   & 0      & 0      & 1/2    \\
O    & 4c   & 0.9722 & 1/4    & 0.5946 \\
O    & 8d   & 0.2998 & 0.0461 & 0.3037 \\ [1ex]
\end{tabular*}

\begin{tabular*}{0.95\columnwidth}{@{\extracolsep{\fill}}ll}
\hline\hline
$R3c$-G & $a$~=~$b$~=~5.559~\AA \;\; $c$~=~13.782~\AA \\
        &  $\alpha$~=~$\beta$~=~90$^\circ$ \;\;
        $\gamma$~=~120$^{\circ}$ \\ [1ex]
\hline
\end{tabular*}
\begin{tabular*}{0.95\columnwidth}{@{\extracolsep{\fill}}ccccc}
Atom & Wyc. & $x$    & $y$    & $z$    \\
Bi   & 6a   & 0      & 0      & 0.0000 \\
Fe   & 6a   & 0      & 0      & 0.7236 \\
O    & 18b  & 0.3156 & 0.2294 & 0.1238 \\ [1ex]
\end{tabular*}

\begin{tabular*}{0.95\columnwidth}{@{\extracolsep{\fill}}ll}
\hline\hline
$Pna2_{1}$-G        & $a$~=~5.702~\AA \;\; $b$~=~5.507~\AA \;\; $c$~=~8.036~\AA \\
(PBE+{\sl U})      & $\alpha$~=~$\beta$~=~$\gamma$~=~90$^{\circ}$ \\  [1ex]
\hline
\end{tabular*}
\begin{tabular*}{0.95\columnwidth}{@{\extracolsep{\fill}}ccccc}
Atom & Wyc. & $x$    & $y$    & $z$    \\
Bi   & 4a   & 0.4435 & 0.0016 & 0.2194 \\
Fe   & 4a   & 0.5015 & 0.5007 & 0.4943 \\
O    & 4a   & 0.2137 & 0.7074 & 0.0519 \\
O    & 4a   & 0.1848 & 0.6876 & 0.4796 \\
O    & 4a   & 0.5302 & 0.4171 & 0.2532 \\ [1ex]
\hline\hline
\end{tabular*}

\end{table}

In addition, we found a number of phases that involve a large stretching of
the ideal cubic cell along the $z$ direction, with $c/a$ aspect ratios
approaching 1.3. In the following we will generically refer to such structures
as {\sl super-tetragonal} or {\sl T} phases. They all favor the C-AFM order
(see Fig.~1), the parallel spin alignment occurring along the stretched
lattice vector. The magnetic interactions along $z$ are weak, as evidenced by
an energy splitting of about 5~meV/f.u. between the C- and G-AFM orders;
accordingly, the ordering temperatures will be relatively low. Three of these
phases are monoclinic ($Cc$-C, $Cm$-C, and $Pc$-C) and one is orthorhombic
($Pna2_{1}$-C); all of them are ferroelectric with a very large polarization
component along [001] (see computed values in Table~II).

More specifically, the $Cc$-C phase presents a polarization in the
(1$\bar{1}$0) plane, as well as relatively small AFD distortions. This type of
monoclinic phase is usually termed $M_{A}$;\cite{vanderbilt01} a similar phase
has been studied theoretically in connection with the super-tetragonal
structures observed experimentally in BFO
films.\cite{bea09,zeches09,wojdel10,hatt10,fn:oldmonoclinic}

The $Pc$-C phase is very similar to $Cc$-C as regards the polar
distortion (thus, it is also $M_{A}$), but it displays a different
O$_6$-rotation pattern.

The $Cm$-C phase displays a polarization in
the (100) plane, and the cell is significantly distorted in the $xy$
plane; such a monoclinic phase is termed $M_{C}$.

The $Pna2_{1}$-C phase is very similar to the $Pna2_{1}$-G structure
discussed above, the stretching of the cell and development of
polarization coinciding with the axis of the in-phase rotations.

Note that all these {\sl T} phases can be viewed as distorted versions of the
ideal super-tetragonal $P4mm$-C structure listed in Table~I. Interestingly, we
found that this $P4mm$-C phase, which is the ground state of
BiCoO$_3$,\cite{azuma08} is a saddle point in BFO's energy landscape.

Our results thus reveal an intricate energy landscape, especially
regarding structures with a large $c/a$ ratio. In this sense, it is
interesting to note that some of the phases reported here are small
distortions of higher-symmetry structures. For example, the $Cm$-C
phase can be shown to be a $Pm$-C structure distorted by the
$M_{3}^{+}$-[$0,y,0$] mode listed in Table~I; by moving from the
$Pm$-C saddle point to the $Cm$-C minimum, BFO gains about
1~meV/f.u. Similarly, the reported $Pc$-C phase is connected with a
higher-symmetry $Cm$-C structure {\sl via} a $M_{3}^{+}$-[$0,0,z$]
distortion.\cite{fn:quantum} Given BFO's manifest complexity, we tend
to view the phases of Table~I as a probably-incomplete list, just
indicative of the rich variety of stable structures that this compound
can present.

Finally, let us stress we explicitly checked that all the above phases
are local minima of the energy, a fact that is remarkable since some
of them (e.g., the pairs formed by $Pnma$-G and $Pna2_{1}$-G, or
$Cc$-C and $Pc$-C) are rather close structurally. It is also
interesting to note that monoclinic phases with such small primitive
cells may be energy minima by themselves, i.e., without the need of
any stabilizing (electric, stress) fields. Note that, to the best of
our knowledge, monoclinic phases in bulk perovskite oxides tend to be
associated with complex solid solutions or large unit cells.  Examples
of the former are the monoclinic $M_{A}$ phase that occurs in
prototype piezoelectric PbZr$_{1-x}$Ti$_{x}$O$_3$,\cite{noheda99} and
the monoclinic $M_{C}$ phase of relaxor
PbZn$_{1/3}$Nb$_{2/3}$O$_3$-PbTiO$_3$.\cite{noheda01} Examples of the
latter occur in BiMnO$_3$ and BiScO$_3$; see the discussion in
Ref.~\onlinecite{haumont09}. It was thus unexpected to discover that
bulk BFO presents such a collection of {\em simple} low-symmetry
minima of the energy.

\subsection{Energy differences between phases}

The relative stability of the phases discussed above is quantified by the
energy differences between them. Disturbingly, Table~I shows that such energy
differences are strongly dependent on the DFT functional used to compute them.
By switching functional we obtained changes in relative stability -- e.g.,
$Pnma$-G is more stable than the {\sl T} phases according to PBEsol+{\sl U}
and LDA+{\sl U}, but less stable according to PBE+{\sl U} -- and even the loss
of stability of one phase -- i.e., the $Pna2_1$-G phase is stable for PBE+{\sl
  U}, but the relaxation of this structure with PBEsol+{\sl U} and LDA+{\sl U}
leads to the $Pnma$-G solution. Thus, we have to ask ourselves: Does any of
these functionals provide an accurate picture of the relative stability of
BFO's phases? Noting that PBEsol is generally more accurate regarding the
structural description of individual phases, can we just rely on the
PBEsol+{\sl U} results?

One would like to address this issue by resorting to a higher-level
first-principles theory. However, performing quantum Monte Carlo
calculations, which are the reference for accuracy in this context, is
well beyond the scope of this work. Simpler schemes like the so-called
hybrid functionals, which are usually considered to be more accurate
than DFT for insulators like BFO, are not well tested for quantifying
relative stabilities in cases like this one. Moreover, structural
predictions with hybrids have been shown to depend strongly on the
underlying generalized gradient approximation,\cite{bilc08} which
invalidates them for the present purposes.

Nevertheless, we were able to make a couple of meaningful comparisons with
experiment. First, we studied the transition between the $R3c$-G and $Pnma$-G
phases that is known to occur under hydrostatic pressure.\cite{fn:hydrostatic}
We obtained (see Fig.~4) transition pressures of about 2~GPa for LDA+{\sl U},
3~GPa for PBEsol+{\sl U}, and 5~GPa for PBE+{\sl U}. Room-temperature
experiments by Haumont {\sl et al}.\cite{haumont09} showed that at 3.5~GPa the
$R3c$-G phase transforms into a monoclinic $C2/m$ structure with a large cell
(made of 12 formula units), and that a second transition at 10~GPa leads to
the $Pnma$-G phase. These results suggest that the $R3c$-G and $Pnma$-G phases
revert their relative stability at a pressure between 3.5~GPa and 10~GPa, a
bracket that can be shifted to 5--14~GPa if the transition lines are
extrapolated to 0~K.\cite{catalan09} Thus, this comparison seems to indicate
that the PBE+{\sl U} is the most accurate theory for relative stability
calculations, and that the LDA+{\sl U} should not be used for these
purposes. We have reached similar conclusions in our work with
Bi$_{1-x}$La$_{x}$FeO$_{3}$ solid solutions;\cite{gonzalez-unp} in that case,
the LDA+{\sl U} predicts a $R3c$-to-$Pnma$ transition for a La content that is
clearly too small to be compatible with the experiments.

Second, we computed the relative stabilities of these phases as a
function of an epitaxial strain corresponding to a square substrate in
the (001) plane, so as to determine the lattice mismatch needed to
stabilize the large-$c/a$ structures.\cite{fn:epitaxial} As shown in
Fig.~5, we obtained strain values of $-$2.3\%, $-$4.0\%, and $-$4.5\%
for PBE+{\sl U}, PBEsol+{\sl U}, and LDA+{\sl U},
respectively. Experimentally it is known that a BFO-(001) thin film
grown on SrTiO$_3$ ($-$1.5\% misfit strain) displays a monoclinic
structure that is an epitaxially-distorted version of the $R3c$ phase
(such a phase is believed to be monoclinic $M_{A}$ with the $Cc$ space
group\cite{daumont10}); we will denote this phase by {\sl R} in the
following. In contrast, when LaAlO$_3$ substrates ($-$4.8\% misfit
strain) are used, a super-tetragonal {\sl T} phase whose symmetry
remains unclear,\cite{bea09} or a co-existence of the {\sl R} and {\sl
  T} phases,\cite{zeches09} has been observed. These results suggest
that the energies of the {\sl R} and {\sl T} phases cross at an
epitaxial compression close to $-$4.8\%. Hence, according to this
criterion, and assuming that our large-$c/a$ phases are good
candidates to be the observed {\sl T} phase, the PBE+{\sl U} curves
would be the least reliable ones. We have reached similar conclusions
in our work with BiFe$_{1-x}$Co$_{x}$O$_{3}$ solid
solutions,\cite{dieguez-unp} where PBE+{\sl U} predicts an {\sl
  R}-to-{\sl T} transition for a Co content that is too small to be
compatible with experiment. Further, these observations seem
consistent with a well-known failure of the PBE approximation: it
tends to render too large tetragonal distortions in ferroelectric
perovskites.\cite{bilc08,wu06}

\begin{figure}
\centering
\includegraphics[width=75mm]{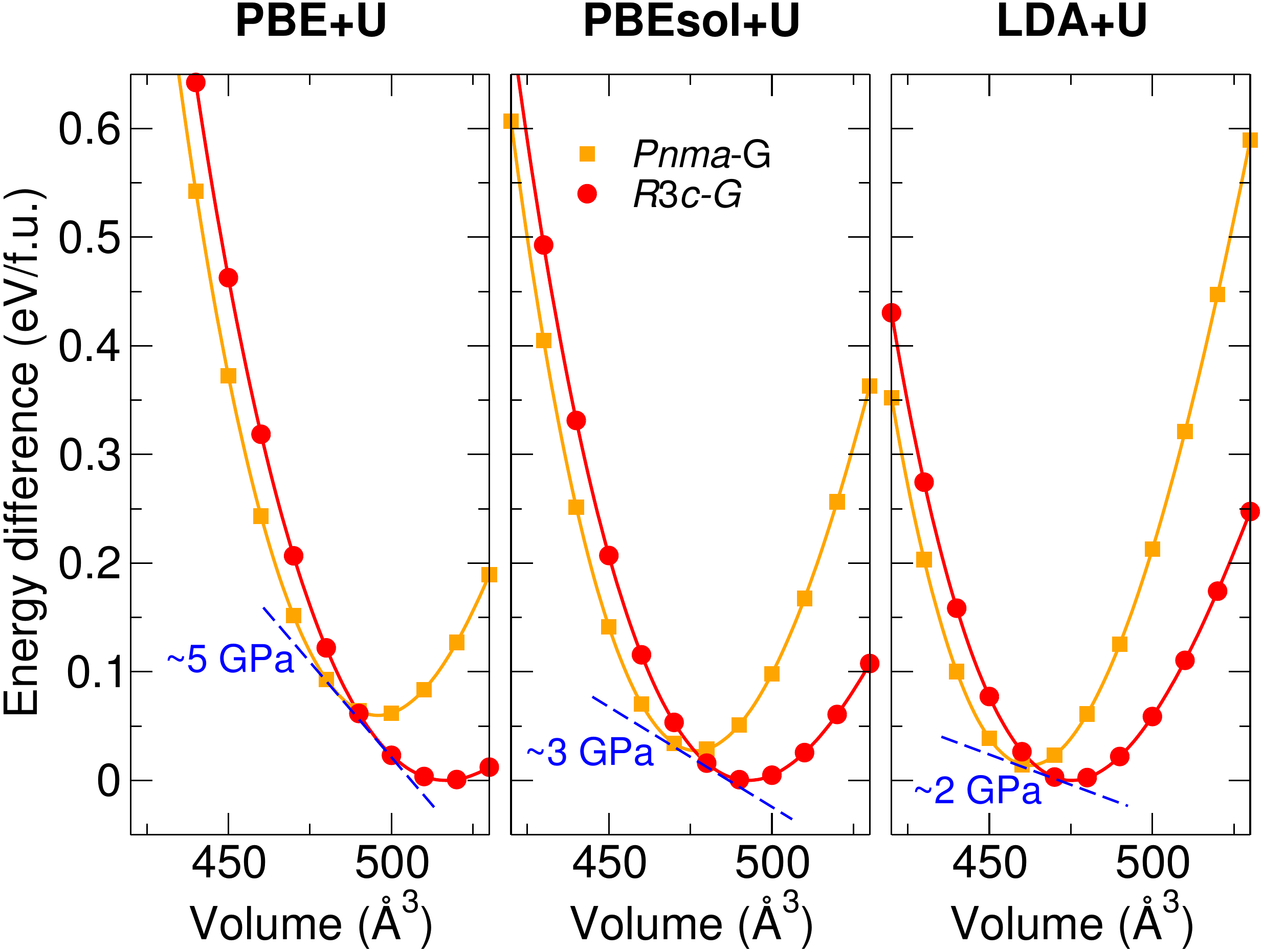}
\caption{(Color online.) Energy versus volume curves for the most stable
  phases of BFO. The
  labels at the top indicate the DFT functional used. The transition
  pressures mentioned in the text were obtained by computing the slope
  of the common tangent of the $R3c$-G and $Pnma$-G curves.}
\label{fig_4}
\end{figure}

In conclusion, while the PBE+{\sl U} and LDA+{\sl U} approaches seem to be
rather accurate in some cases, they also render clearly wrong predictions in
others. In this respect, PBEsol+{\sl U} seems to be a reasonable compromise,
as it constitutes the overall most accurate DFT theory available to
us. Nevertheless, because PBE+{\sl U} performs well as regards the relative
stability of the $R3c$-G and $Pnma$-G phases, we believe that the PBE+{\sl U}
prediction of the new ferroelectric phase $Pna2_{1}$-G, structurally very
similar to $Pnma$-G, deserves some attention. Finally, let us note that the
choice of $U$ also has en effect on the energy differences of Table~I. Yet,
for {\sl U} values in the 3--5~eV range, such effects are small as compared
with the ones we have discussed.

\section{Discussion}

Our results have direct implications for current experimental work on
the structural characterization and phase transitions of BFO,
especially regarding the epitaxially compressed films in which
super-tetragonal phases were discovered. Further, they also provide us
with information that is relevant to the effective modeling of BFO's
structural transitions, at both the macroscopic (Landau-type theories)
and atomistic (effective Hamiltonians) levels. In the following we
discuss all these aspects. To conclude this Section, we comment on
Bi's ability to form very different and stable {\em coordination
  complexes} with oxygen, as this seems to be the factor responsible
for the observed structural richness of BFO.

\begin{figure}
\centering
\includegraphics[width=75mm]{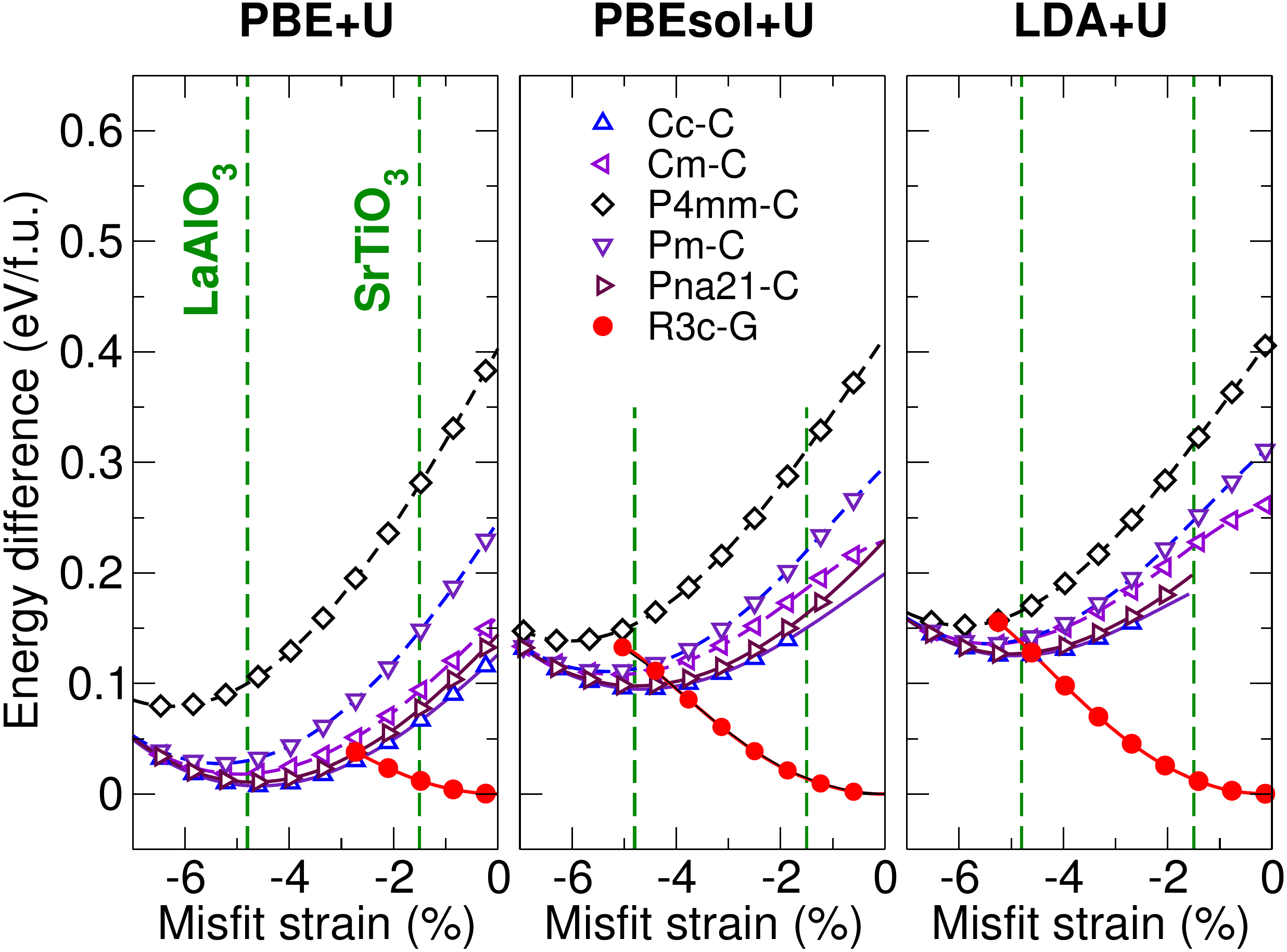}
\caption{(Color online.) Energy of various BFO phases as a function of the
  misfit (epitaxial) strain corresponding to a square (001)-oriented
  substrate. The labels at the top indicate the DFT functional used. Note that
  the $R3c$-G phase reduces its symmetry to $Cc$-G in these epitaxial
  conditions.}
\label{fig_5}
\end{figure}

\subsection{Implications for experimental work}

\subsubsection{{\em Super-tetragonal} phases in BiFeO$_3$ films}

The recent works by B\'ea {\sl et al}.\cite{bea09} and Zeches {\sl et
  al}.\cite{zeches09} have shown that it is possible to obtain a novel
phase of BFO if thin films are grown on strongly compressive
substrates like LaAlO$_3$-(001). Experimentally, this {\sl T} phase
presents a very large $c/a$ ratio of about 1.23, and an out-of-plane
polarization $P_{z}\approx$~0.8~C/m$^2$. First-principles
studies\cite{zeches09,wojdel10,hatt10} have identified the {\sl T}
phase with a monoclinic $Cc$ structure for which LDA+{\sl U}
calculations predict $c/a\approx$~1.23 and
$P_{z}\approx$~1.5~C/m$^2$. Thus, there is a large quantitative
discrepancy between theory and experiment as regards the value of
$P_{z}$, which suggests that the identification of the simulated and
experimental phases may be incorrect.

Our present results show that there are many possible {\sl T} phases -- e.g.,
the low-energy $Pc$-C, $Cm$-C, $Pna2_1$-C, and $Cc$-C structures that we found
-- that might correspond to the one experimentally realized in the BFO
films. Indeed, as shown in Fig.~5, all our large-$c/a$ phases are essentially
degenerate in energy for values of the epitaxial strain corresponding to a
LaAlO$_3$ substrate. Moreover, at the PBEsol+{\sl U} level -- which we have
adopted as the DFT flavor of choice for BFO --, all these phases have their
energy minimum at a misfit strain of about $-$4.8\%, implying that any of them
can form a stable BFO film under such epitaxial conditions.

Because our {\sl T} phases are an almost perfect epitaxial match with
the LaAlO$_3$ substrate, the structural and polarization data in
Tables II and III can be compared with the experimental results
directly. Most remarkably, our results show that phases with very
similar $c/a$ ratios can display rather different polarization
values. Indeed, the $Pc$-C phase (with a $c/a$ of 1.27) presents
$P_{z}\approx$~1.1~C/m$^2$, while the $Cm$-C and $Pna2_1$-C phases
(with $c/a$'s of 1.26 and 1.25, respectively) present
$P_{z}\approx$~1.4~C/m$^2$. Hence, our $Pc$-C structure seems to be
the best candidate to represent the {\sl T} phase realized in the BFO
films investigated experimentally; the quantitative disagreement
between the measured and predicted $P_{z}$'s would be below 40\%, a
clear improvement upon the previously reported 90\% difference.

Let us also note that, because our {\sl T} phases are so close in
energy, the question of which one is realized experimentally may
depend on subtle details not considered in this work. Thus, for
example, two of these phases ($Pc$-C and $Cm$-C) present no {\em
  tilts} (i.e., rotations around the [100] and [010] axes) of the
O$_6$ octahedra, which may make them preferable if the BFO films are
grown on (001) substrates that clamp such distortions
strongly. Similarly, a rectangular substrate might favor the $Cm$-C
phase, whose cell tends to distort in the $xy$ plane, etc.

Finally, we have very recently become aware of new
results\cite{chen10,chen11,nam10} showing that both $M_{C}$ and
$M_{A}$ monoclinic phases with large-$c/a$ ratios can be realized in
epitaxially compressed BFO-(001) films. Such findings further support
the physical relevance of the present study.

\subsubsection{Structural transitions in bulk BiFeO$_3$}

Our calculations were restricted to the limit of low temperatures, and
do not allow for a conclusive discussion of temperature-driven effects
and transitions in BFO.\cite{fn:thermal,zhong95} Nevertheless, a few
comments can be made based on the obtained (large) energy differences
between some relevant phases. Indeed, our results seem consistent with
experiments\cite{catalan09,palai08,arnold10} showing that, as a
function of increasing temperature, BFO's ferroelectric $R3c$ phase
transforms into an orthorhombic $Pmna$ structure at $T\approx$~1100~K,
to then become cubic $Pm\bar{3}m$ at $T\approx$~1200~K. More
specifically, the PBEsol+{\sl U} results of Table~I show that the
$R3c$-G and $Pnma$-G minima are very close in energy and constitute
strong instabilities of the prototype $Pm\bar{3}m$ structure, which
lies about 900~meV/f.u. above them, as consistent with the fact that
BFO's cubic phase can be observed only at very high
temperatures. Moreover, $R3c$-G and $Pnma$-G constitute BFO's most
stable phases, with a large margin over other structures (e.g., the
ferroelectric $R3m$-G and $Amm2$-G, or paraelectric $R\bar{3}c$-G and
$I4/mcm$-G, listed in Table~I) that are common among perovskite
oxides. Hence, our results seem incompatible with the $R3c \rightarrow
I4/mcm \rightarrow Pm\bar{3}m$ transition sequence obtained by Kornev
{\sl et al}.\cite{kornev07} from Monte Carlo simulations of
first-principles-derived effective Hamiltonians; we found that the
$I4/mcm$ structure has a relatively high energy and is thus unlikely
to occur instead of $Pnma$.

As regards pressure-driven transitions, our results confirm that under
compression BFO's $R3c$-G phase loses stability in favor of the $Pnma$-G
structure.\cite{haumont09,ravindran06} Additionally, it is worth noting that,
at the PBE+{\sl U} level, we found a $Pna2_1$-G phase (see Table~I) whose
stability is also favored by compression and which nearly becomes the ground
state in the pressure range in which $R3c$-G and $Pnma$-G revert their
relative stability (results not shown here). Given that PBE+{\sl U} seems the
most accurate DFT flavor for the description of these pressure-induced
transformations (see Section III.B), it seems wise to bear in mind the
possibility that such a $Pna2_1$-G structure might occur, especially
considering that the nature of the phase intermediate between $R3c$-G and
$Pnma$-G remains unclear.\cite{haumont09}

\subsection{Implications for modeling work}

Our results clearly demonstrate that, in spite of its apparent
simplicity, BiFeO$_3$ is extraordinarily complex from the structural
point of view. In the following sections we will {\em quantify} such a
complexity, adopting the perspective of someone who is interested in
determining the simplest possible model, either macroscopic or
atomistic, that captures accurately BFO's structural diversity. Our
analysis shows that BFO is much more challenging to model than {\em
  traditional} ferroelectric perovskites like BaTiO$_3$, PbTiO$_3$, or
even PbZr$_{1-x}$Ti$_{x}$O$_3$.

\subsubsection{Primary and secondary distortions in BiFeO$_3$}

By analyzing the BFO phases described in Table~I, it is possible to identify
three primary distortion types (or primary order parameters) whose occurrence
can explain all the symmetry reductions of interest and which must be
considered in any theory of BFO's structural phase transitions: A polar
distortion that can in principle be oriented along any spatial direction
($\Gamma_{4}^{-}$ symmetry), and in-phase ($M_{3}^{+}$) and anti-phase
($R_{4}^{+}$) O$_6$ rotations around the three Cartesian axes. The atomic
displacements associated with the two AFD order parameters (i.e., the
oxygen-octahedra rotations) are uniquely defined by symmetry; hence, these
modes are trivial in this sense. In contrast, the polar distortions are not
determined by symmetry: any combination of $\Gamma_{4}^{-}$-like displacements
of the Bi, Fe, and O sub-lattices is in principle valid. Following the usual
first-principles approach to {\em simple} ferroelectric perovskites like
BaTiO$_3$ or PbTiO$_3$,\cite{kingsmith94} one would determine the specific
atomic displacements that define the FE order parameter by computing the
unstable (soft) polar mode of the cubic phase of the compound; the result thus
obtained for BFO is depicted in Fig.~3(a). In materials like BaTiO$_3$, such a
soft mode captures very accurately the atomic distortions associated to the
relevant FE phases, e.g., tetragonal $P4mm$ and rhombohedral $R3m$. It is not
obvious that the same will be true for BFO, where we would like to describe
{\em simultaneously} super-tetragonal phases, which imply a very large
distortion of the cubic cell, and the rhombohedral ground state, where the
polar distortion co-exists with very large O$_6$ rotations. Interestingly, we
were able to demonstrate that the traditional approach works well for BFO: We
performed a mode-by-mode decomposition of the atomic distortions connecting
the prototype $Pm\bar{3}m$-G phase with the $P4mm$-C (as representative of our
large-$c/a$ phases) and $R3c$-G structures, and checked that the
$\Gamma_{4}^{-}$-like component is captured very accurately by the soft FE
mode of the cubic phase (to within a 93\% for $P4mm$-C and 99\% for
$R3c$-G). We can thus conclude that it is possible to describe {\em all} the
FE phases of BFO with relatively simple theories that include only one polar
mode.

The three primary order parameters described above are clearly the
driving force for the structural transitions in BFO. For a given phase
of the material, the occurrence of a particular combination of such
primary distortions involves a specific breaking of the $Pm\bar{3}m$
symmetry of the cubic perovskite structure, which in turn results in
the {\em activation} of secondary order parameters that become allowed
in the low-symmetry phase. The most significant secondary distortions
that we found in our BFO's phases are listed in the last column of
Table~I and sketched in Fig.~3. There is a considerable number of such
secondary modes; the ones involving the largest atomic displacements
can be easily grouped in two categories: AFE patterns (see (c) to (f)
modes in Fig.~3) and twisting modes of the O$_6$ octahedra ((b) in
Fig.~3). In this sense, BFO is very different from ferroelectrics like
BaTiO$_3$ or PbTiO$_3$, where the relevant FE phases do not present
any secondary modes (note the absence of additional distortions for
the $P4mm$, $Amm2$, and $R3m$ symmetries listed in Table~I, which are
the relevant ones for BaTiO$_3$ and PbTiO$_3$). One thus needs to
wonder: How important are these secondary distortions?  Do they play a
role in determining the energetics and relative stability of BFO's
phases, or can they be ignored in an effective theory of BFO's
structural phase transitions?\cite{fn:strain}

We quantified the importance of the secondary modes in the following
approximate manner: We considered the PBEsol+{\sl U} equilibrium structures of
all the relevant phases, artificially set to zero the secondary atomic
distortions, and computed the energy of the modified structures. The obtained
energy increments with respect to the actual equilibrium phases are
very significant: they range from tens of meV/f.u.  for the monoclinic
($Pc$-C, $Cm$-C, and $Cc$-C) phases to more than a hundred for the
orthorhombic ($Pna2_1$-C and $Pnma$-G) ones. A more exact estimate can easily
be performed for $Pnma$-G, as we found that in this case the most relevant
secondary modes are clearly associated to Bi displacements: By fixing the Bi
ions at their high-symmetry positions and relaxing all other structural
parameters, we obtained an energy increase of 125~meV/f.u. with respect to the
fully-relaxed $Pnma$-G structure. Thus, our results show that the energy
changes associated to the secondary modes are of the same magnitude as the
energy differences between different phases, which implies that these modes
play a key role in determining BFO's phase diagram. In particular, the large
effects obtained for the orthorhombic phases indicate that their stability
depends crucially on occurrence of the AFE patterns associated to Bi's
off-centering. We can thus conclude that an effective theory of BFO's
structural transitions must account for the effect of these secondary modes.

\subsubsection{Phenomenological theories}

The Devonshire-Landau phenomenological approach to phase transitions
in bulk ferroelectrics,\cite{devonshire49,iniguez01} and its extension
to epitaxially-constrained films,\cite{pertsev98,dieguez04}
constitutes the simplest, yet powerful, theory that one might try to
use to model BFO. Working out such a theory for BFO -- i.e.,
determining the simplest possible Landau potential and temperature
dependence of the parameters -- constitutes a great challenge that, as
far as we know, remains to be tackled. In the following we discuss
what our results imply as regards the Landau theory of BFO.

In order to describe all the known phases of this compound, the corresponding
Landau potential should be written in terms of a three-dimensional FE
polarization ${\bf P}$ (which would correspond to the atomic distortions
discussed in Section IV.B.1), as well as two AFD order parameters associated,
respectively, to in-phase and anti-phase O$_6$-octahedra rotations. The cross
terms between these three three-dimensional primary order parameters, and the
additional terms that will appear if a non-zero epitaxial strain is
considered, should allow us to reproduce the intricate energy landscape of
BFO and its low-symmetry minima.

Indeed, in cases with several order parameters, it is possible to obtain
stable low-symmetry phases from low-order Landau potentials. Imagine, for
example, a FE perovskite that develops a polarization along the [1,1,1]
direction as well as an in-phase O$_6$ rotation around the [0,0,1] axis. Such
instabilities can be described with a Landau potential truncated at 4th order
in both the FE and AFD order parameters. The resulting phase would have a
monoclinic $Pc$ ($M_{A}$) symmetry, exactly as the $Pc$-C structure of
Table~I. Hence, according to this example, it might be possible to describe
all BFO's phases with a low-order Landau theory.

\begin{figure}
\centering
\subfigure[]{
\includegraphics[width=33mm]{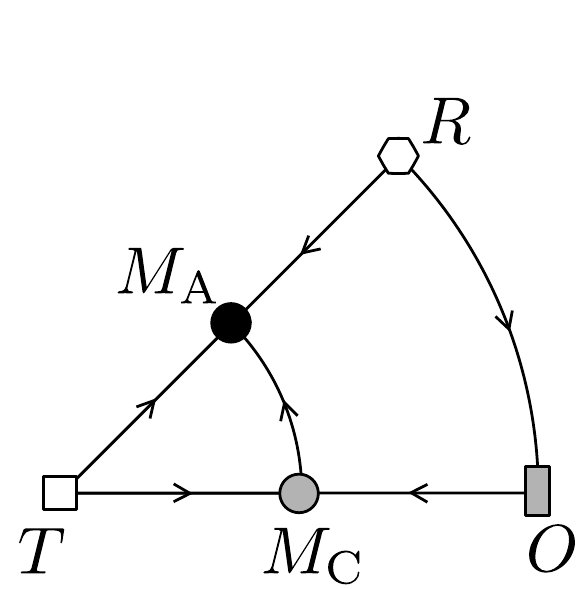}
}
\hspace{5mm}
\subfigure[]{
\includegraphics[width=33mm]{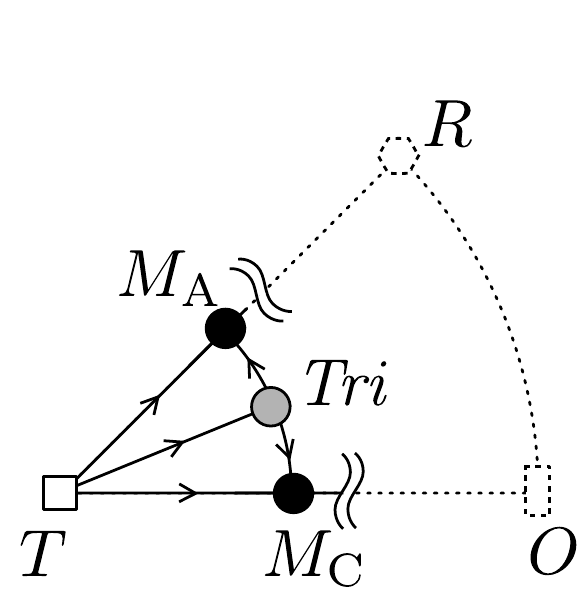}
}
\caption{Energy landscape diagrams as introduced in
  Ref.~\onlinecite{vanderbilt01}. Filled, open, and shaded symbols correspond
  to minima, maxima, and saddle points of the energy, respectively. The {\sl
    T}, {\sl R}, {\sl O}, and {\sl M} labels denote phases with exact
  tetragonal, rhombohedral, orthorhombic, and monoclinic symmetries,
  respectively. (a) Simplest scenario that gives raise to a monoclinic $M_A$
  minimum.\cite{vanderbilt01} (b) Simplest scenario that gives raise to
  simultaneous $M_A$ and $M_C$ minima; our discussion focuses on the left part
  of this diagram (see text).}
\label{fig_6}
\end{figure}

However, our results show that the Landau theory for BFO would be
significantly more complicated, especially in what regards the relative
stability of the large-$c/a$ phases. To illustrate this point, let us consider
a simplified version of BFO in which only FE distortions and cell strains are
allowed, and try to determine the order of the Landau potential $F({\bf P})$
required to describe ferroelectricity in such a system.

In a landmark article, Vanderbilt and Cohen\cite{vanderbilt01} analyzed the
form of the Landau potential needed to describe low-symmetry phases in FE
perovskites. In essence, they showed that a potential $F({\bf P})$ can present
tetragonal or rhombohedral minima if expanded up to 4th order in ${\bf P}$;
the occurrence of orthorhombic minima requires a 6th-order theory, and one
needs to go up 8th order to have minima of monoclinic symmetry. This work was
essential to understand which Landau potentials are needed to describe the
monoclinic phases that were being found at the time in perovskite solid
solutions such as PbZr$_{1-x}$Ti$_{x}$O$_3$ ($M_{A}$ type)\cite{noheda99} and
PbZn$_{1/3}$Nb$_{2/3}$O$_3$-PbTiO$_3$ ($M_{C}$ type).\cite{noheda01} The
energy landscape associated to an 8th-order potential with a monoclinic
$M_{A}$ minimum is sketched in Fig.~6(a), following the convenient
representation scheme introduced in Ref.~\onlinecite{vanderbilt01}.

We simulated our simplified (FE-only) version of BFO by forcing the material
to have a 5-atom unit cell in which only polar ($\Gamma_{4}^{-}$) distortions
and cell strains are allowed. (Of course, this cell was appropriately doubled
to capture the G- and C-AFM spin arrangements.) If we impose such a constraint
to the phases in Table~I, we immediately recover the symmetries that were
broken by the AFD modes: The $Pc$-C and $Cc$-C phases reduce to a single
monoclinic $M_{A}$ structure with space group $Cm$-C; the $Cm$-C phase changes
to a monoclinic $M_{C}$ with $Pm$-C symmetry; $R3c$-G gives us a $R3m$-G phase
analogous to BaTiO$_3$'s ground state, etc. We can then consider two
additional phases -- namely, the super-tetragonal $P4mm$-C and orthorhombic
$Amm2$-G listed in Table~I --, to sketch the energy landscape of
Fig.~6(b). (To plot Fig.~6(b), the structural stability against $\Gamma$-like
distortions of the {\sl T} and {\sl M} phases was explicitly checked. We have
divided the diagram in two sectors to emphasize that the distortions
connecting the super-tetragonal phases with the rhombohedral and orthorhombic
structures are very large.)  The most notable feature of this energy diagram
is that it presents two inequivalent monoclinic minima, as opposed to only one
as in Fig.~6(a). Further, if we follow the lowest-energy path connecting the
$M_{A}$ and $M_{C}$ minima through triclinic structures, we will necessarily
cross either a saddle point (case depicted in Fig.~6(b)) or a
maximum. According to the analysis of Ref.~\onlinecite{vanderbilt01}, the
existence of a triclinic saddle point requires a Landau potential of 10th
order, while a 12th-order theory is needed to have a triclinic maximum. Note
that Landau potentials of such a high order are unheard of among FE
perovskites, even if complex solid solutions are considered. Amusingly, in
their paper\cite{vanderbilt01} Vanderbilt and Cohen justified the interest of
discussing theories of very high order by writing that ``the discovery (or
synthesis) of a material having such a behavior may be challenging, but is by
no means impossible.''  Our analysis shows that BFO (even a simplified version
of it) is such a material.\cite{fn:lower-order}

\subsubsection{Atomistic theories}

Effective theories of the inter-atomic interactions in ferroelectric
perovskites, with parameters computed from first-principles, were
introduced in the early 90's by Rabe and Vanderbilt.\cite{zhong94}
Ever since, these so-called {\em effective Hamiltonians} have made it
possible to perform statistical simulations of increasingly complex
materials, from crystalline BaTiO$_3$\cite{zhong94} to disordered
PbZr$_{1-x}$Ti$_{x}$O$_{3}$,\cite{bellaiche00} successfully
reproducing temperature-driven phase transitions, response properties,
etc. More recently, an effective Hamiltonian for BFO has been derived
by Kornev {\sl et al}.,\cite{kornev07} who thus extended the approach
to incorporate magneto-structural interactions in the model. Such a
groundbreaking development has led to great physical insight into
BFO's ferroelectric and magnetoelectric
properties,\cite{kornev07,albrecht10} as well as into the material's
behavior under applied electric\cite{lisenkov09a} and
magnetic\cite{lisenkov09b} fields.  On the other hand, in view of
recent experimental results, we now know that some of the model
predictions (e.g., the occurrence of a $I4/mcm$ phase at high
temperature) are questionable. In the following we briefly summarize
what our results teach us about how to construct an accurate effective
Hamiltonian for BFO, extracting the corresponding conclusions as
regards the theory of Kornev {\sl et al}.

The first step of the classic approach to constructing effective Hamiltonians
consists in identifying the relevant {\em local} distortions that must be
retained in the model, so that we can use a coarse-grained representation of
the atoms in the unit cell of our compound. In the case of BFO, there are
clearly two local distortions that need to be considered: (1) a polar
displacement pattern compatible with the FE ($\Gamma_{4}^{-}$) soft mode of
Fig.~3(a), and (2) the rotation of individual O$_6$ octahedra around an
arbitrary axes, whose in-phase ($M_{3}^{+}$) and anti-phase ($R_{4}^{+}$)
repetition throughout the crystal reproduces the relevant AFD modes. As shown
in Section~IV.B.1, it is enough to consider one local polar mode to reproduce
the FE distortion of the $R3c$-G ground state and large-$c/a$ phases, which
allows us to work with a relatively simple model. A first-principles effective
Hamiltonian considering these two types of local variables was first
constructed to study SrTiO$_3$,\cite{zhong95} and this was also the starting
point of the work of Kornev {\sl et al}. for BFO.

In Section~IV.B.1 we demonstrated the importance of the secondary
distortions in determining the relative stability of BFO's phases. The
most relevant secondary modes are clearly the Bi-related AFE patterns
that occur in the $Pnma$-G and $Pna2_1$-C phases. Fortunately, it is
possible to incorporate such effects in an effective Hamiltonian
without extending or complicating the model: We can choose the above
mentioned local polar modes to be centered at the Bi atoms, as
sketched in Fig.~7(a), so that (i) their homogeneous repetition
throughout the crystal reproduces the FE soft mode of Fig.~3(a) and
(ii) the zone-boundary modulations reproduce approximately the most
relevant AFE distortions of the Bi atoms. Note that, alternatively,
one could think of using local polar modes centered at the Fe atoms
(see Fig.~7(b)). However, while this option is valid to reproduce
BFO's FE distortions, it fails to capture Bi's AFE patterns (a
zone-boundary modulation of the Fe-centered modes results in null Bi
displacements).\cite{fn:localmodes} Consequently, an effective model
based on Fe-centered modes will put a considerable energy penalty on
the $Pnma$-G and similar phases. Such was the approach adopted by
Kornev {\sl et al}., which may explain their prediction that a
$I4/mcm$ structure, and not $Pnma$, occurs in the phase diagram of
bulk BFO.

\begin{figure}
\centering
\subfigure[]{
\includegraphics[width=37mm]{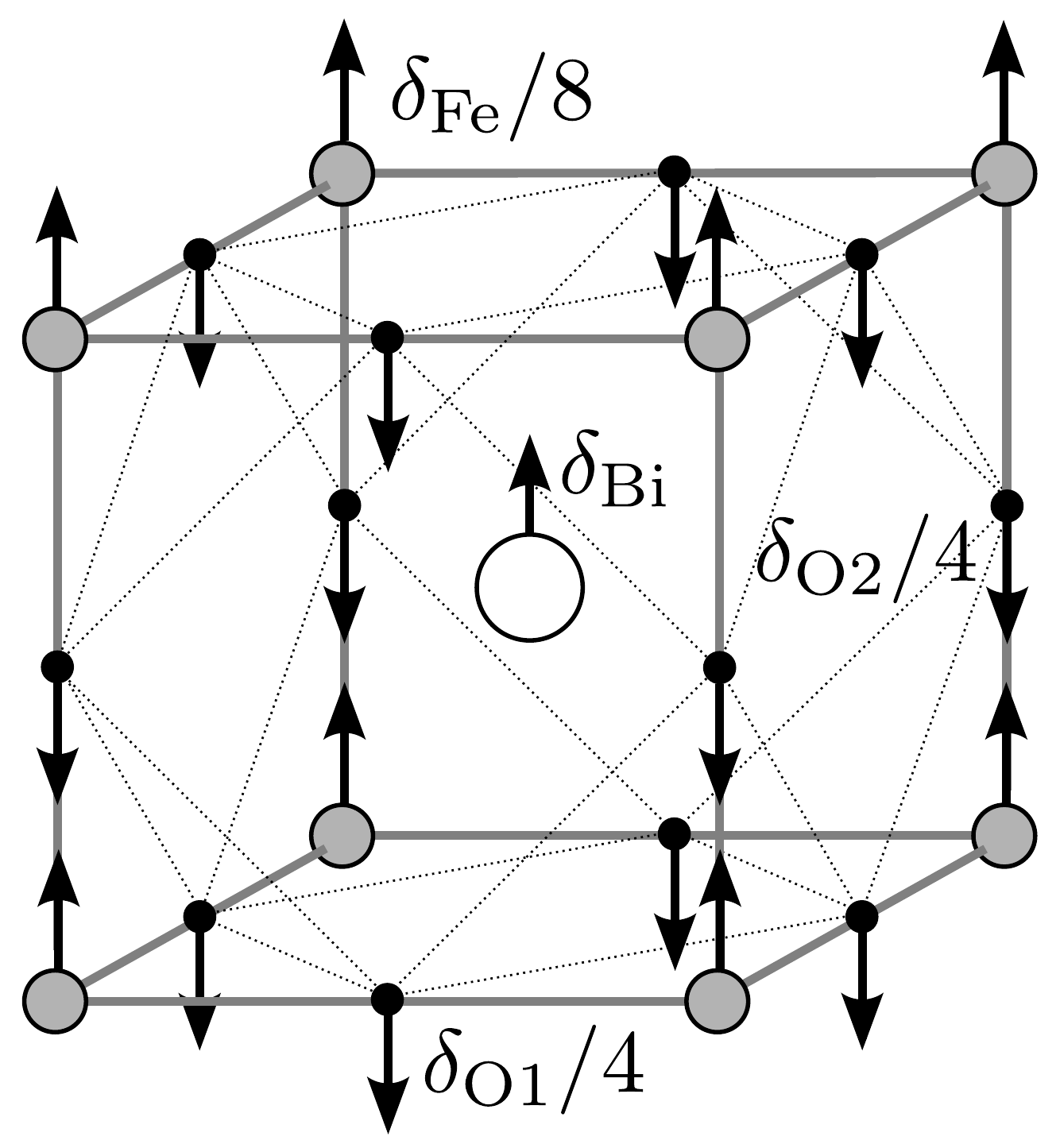}
}
\hspace{4mm}
\subfigure[]{
\includegraphics[width=37mm]{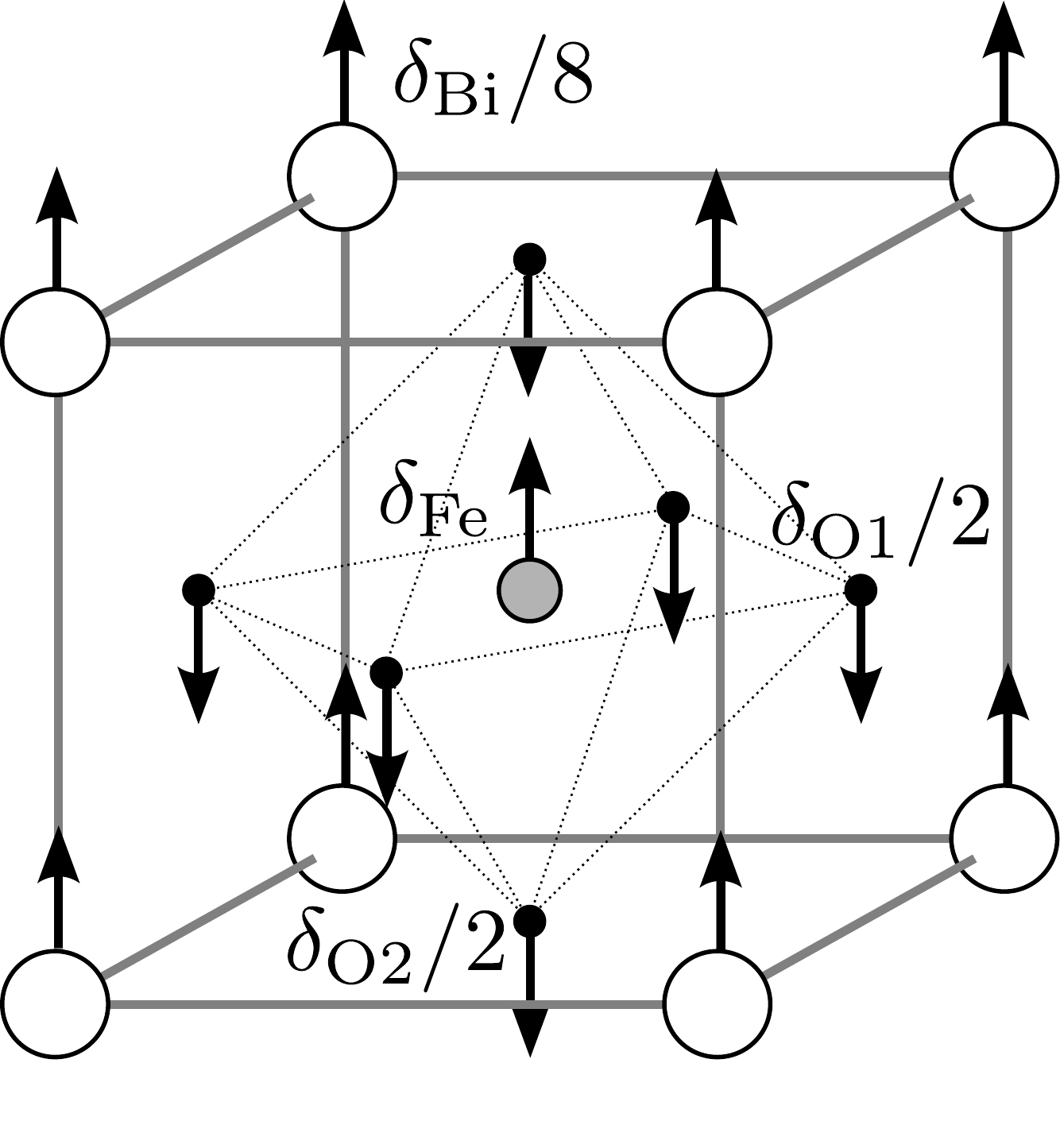}
}
\caption{Examples of local polar modes that can be used as variables of an
  effective Hamiltonian for BFO.  (a) Centered at the Bi atom. (b) Centered at
  the Fe atom.  The quantities $\delta_I$ identify displacements of the
  corresponding $I$ atom in the soft FE mode of the system; they should be
  divided by a factor that takes into account how many cells share atom $I$.
  For BFO we obtained $\delta_{\rm Bi} = 0.80$, $\delta_{\rm Fe} = 0.06$,
  $\delta_{\rm O1} = -0.42$, and $\delta_{\rm O2} = -0.04$.  }
\label{fig_7}
\end{figure}

As regards the rest of (less important) secondary modes, it might be
possible to incorporate their effect by suitably {\em renormalizing}
the Hamiltonian parameters. To make this idea more precise, let us
denote by ${\boldsymbol u}$ (resp. ${\boldsymbol v}$) the distortions
that will (resp. will not) be explicitly considered in the model. The
usual effective Hamiltonians $H_{\rm eff}({\boldsymbol u})$, which
work well for materials like BaTiO$_3$ in which secondary distortions
are clearly not critical, can be formally defined as:
\begin{equation}
  H_{\rm
    eff}({\boldsymbol u}) \; \approx \; E({\boldsymbol u},{\boldsymbol
    v})|_{{\boldsymbol v}=0} \, ,
\end{equation}
where $E({\boldsymbol u},{\boldsymbol v})$ is the first-principles energy of
an arbitrary configuration of the compound. In contrast, we could define an
effective Hamiltonian $\tilde{H}_{\rm eff}({\boldsymbol u})$ designed to
account for the effect of secondary distortions as:
\begin{equation}
  \tilde{H}_{\rm
    eff}({\boldsymbol u}) \; \approx \; min_{{\boldsymbol v}} \,
  E({\boldsymbol u},{\boldsymbol v}) \, .
\end{equation}
Such a refined approach should improve the accuracy of the models in all
cases, and it might prove critical to obtain correct results for compounds as
challenging as BFO. The implementation of these ideas remains for future work.

\begin{figure}
\centering
\subfigure[~$R3c$-G]{
\includegraphics[height=80mm, angle=-90]{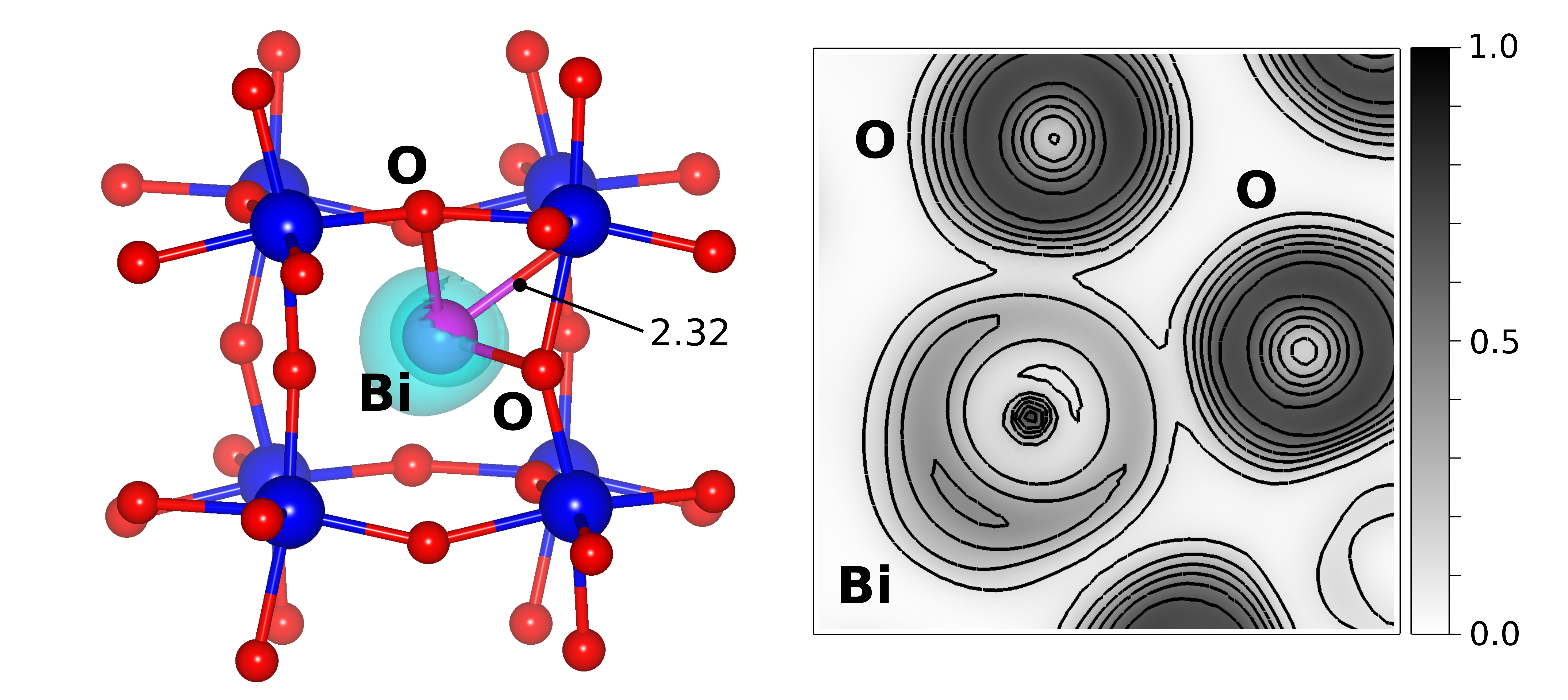}
}
\hspace{4mm}
\subfigure[~$Cc$-C]{
\includegraphics[height=80mm, angle=-90]{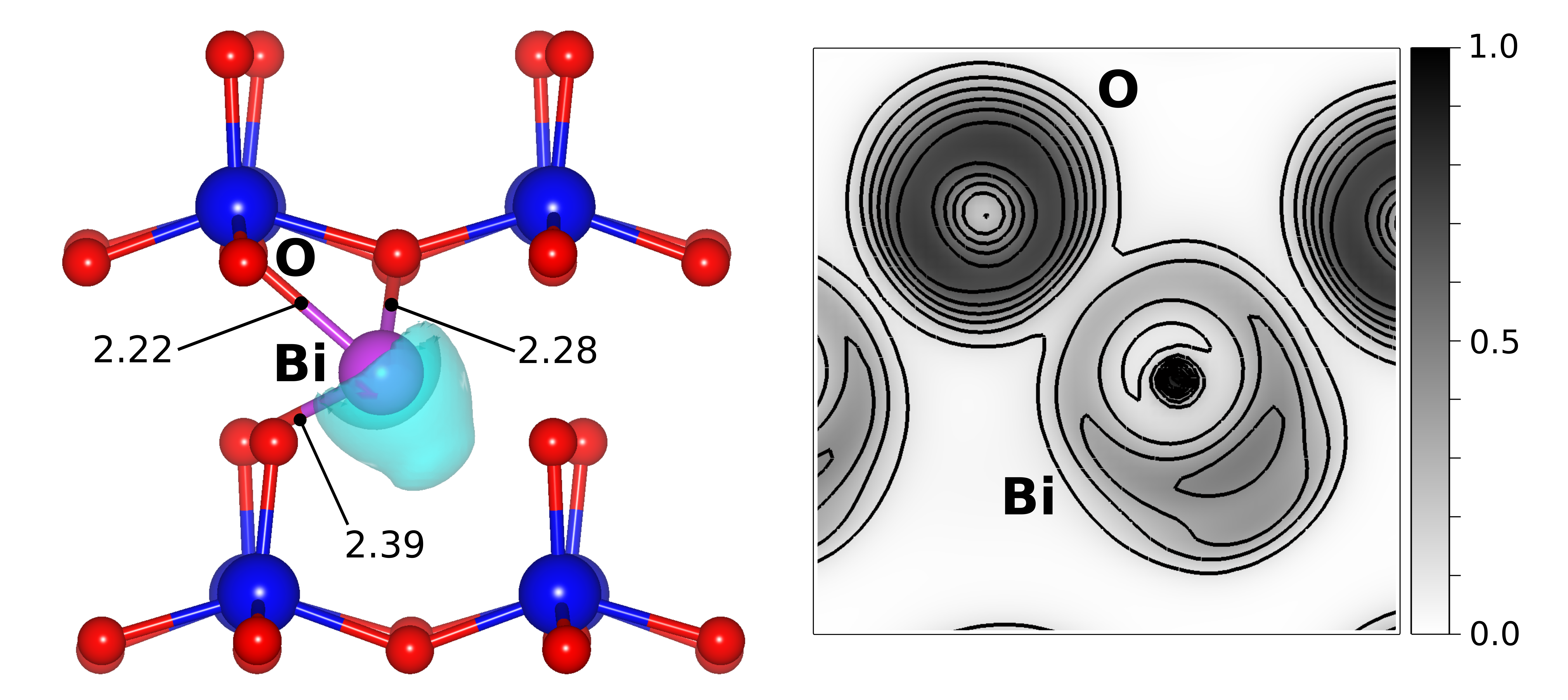}
}
\hspace{4mm}
\subfigure[~$Pnma$-G]{
\includegraphics[height=80mm, angle=-90]{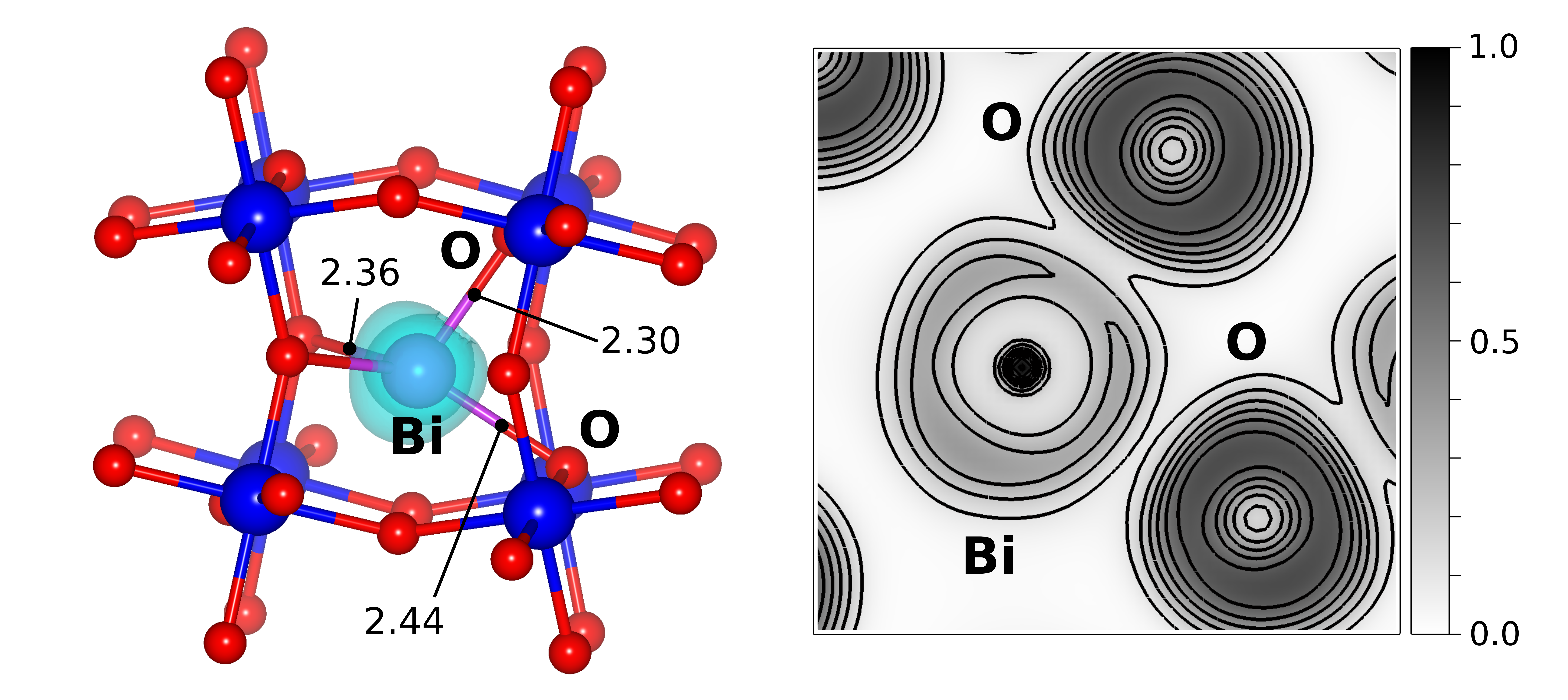}
}
\caption{(Color online.) Electronic-localization-function (ELF) maps computed
  for the $R3c$-G, $Cc$-C, and $Pnma$-G phases. The figures on the left show
  the isosurface for an ELF value of 0.3 superimposed to the atomic structure;
  on the right we show the ELF contour plots in the planes defined by the
  labeled ions. We also indicate the shortest Bi--O distances (in Angstrom) as
  obtained from PBEsol+{\sl U} calculations.}
\label{fig_8}
\end{figure}

\begin{figure}
\centering
\subfigure[~$R3c$-G]{
\includegraphics[height=32mm]{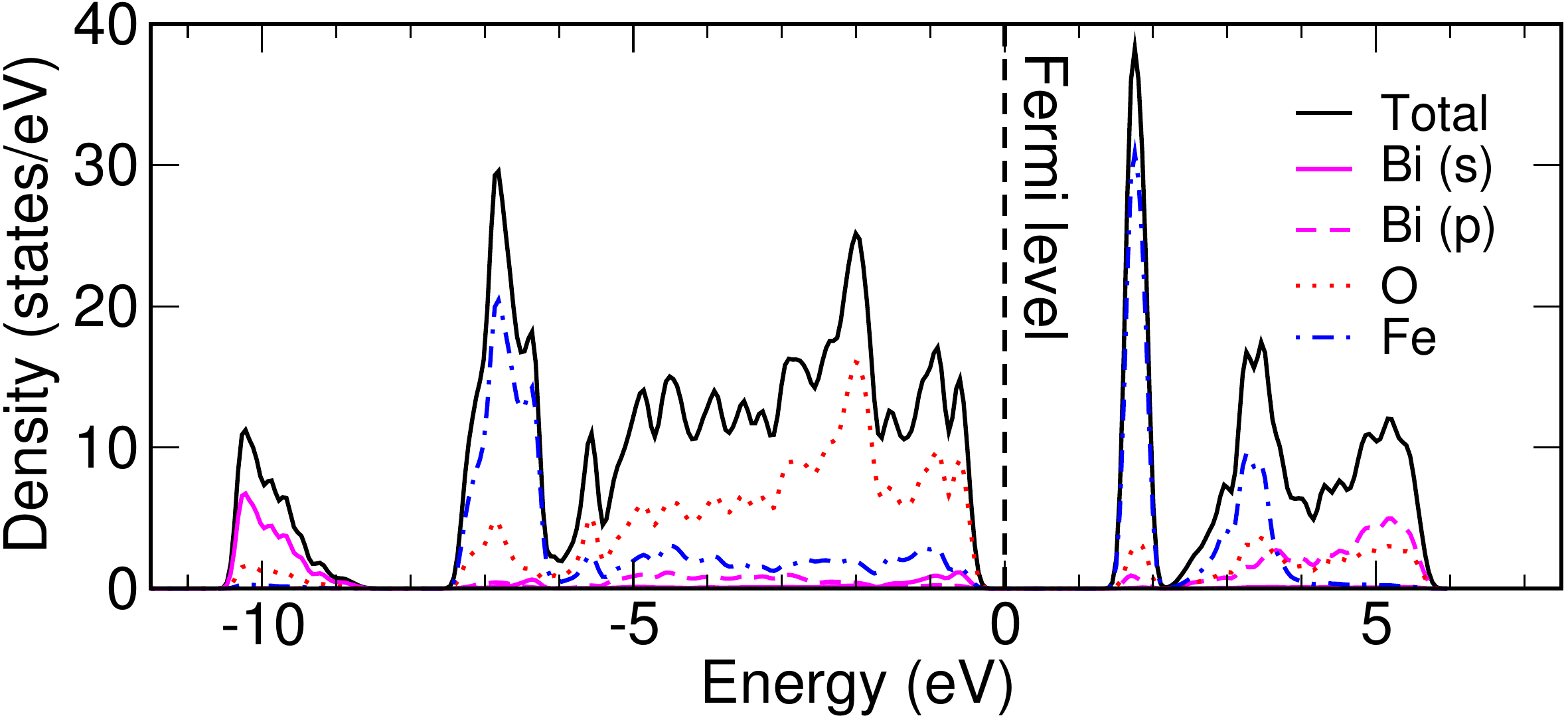}
}
\hspace{4mm}
\subfigure[~$Cc$-C]{
\includegraphics[height=32mm]{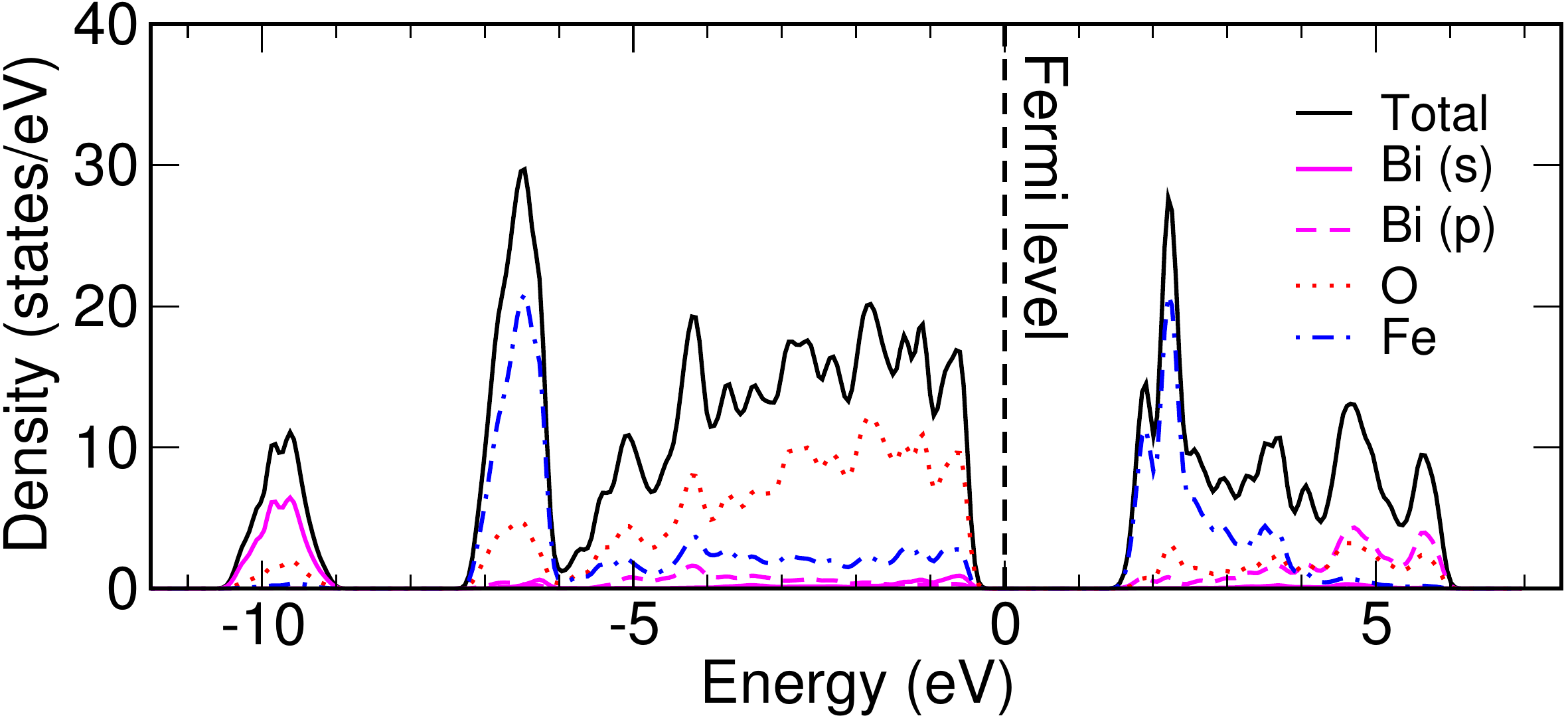}
}
\hspace{4mm}
\subfigure[~$Pnma$-G]{
\includegraphics[height=32mm]{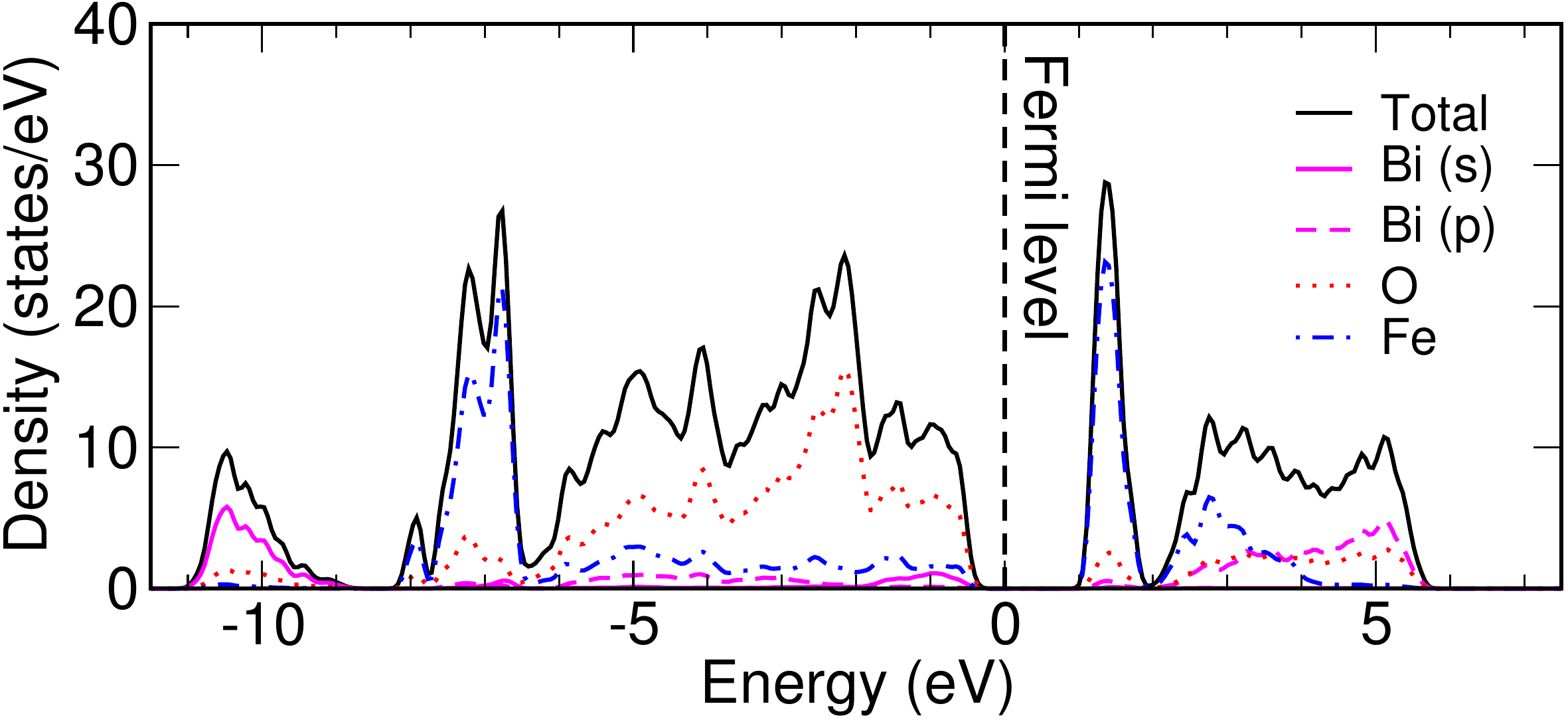}
}
\caption{(Color online.) Electronic density of states for the $R3c$-G, $Cc$-C,
  and $Pnma$-G phases, as obtained from PBEsol+{\sl U} calculations. Note
  that, in these AFM phases, the results for the spin-up and spin-down channels
  are identical.}
\end{figure}

\subsection{The role of Bismuth}

The Bi cations play a key role in BFO's structural transitions. This can be
predicted already from very simple steric arguments: In Bi{\sl M}O$_3$
perovskites, where {\sl M} is a first-row transition metal, the lattice
parameter is essentially determined by the ionic radii of the metal and oxygen
ions. This situation, which corresponds to a small value of the so-called {\sl
  tolerance factor},\cite{iniguez03} tends to result in either the
off-centering of the Bi$^{3+}$ cation or the occurrence of AFD modes, both of
which imply the shortening of some Bi--O bonds.\cite{fn:shannon} This is
exactly what is commonly observed in Bi{\sl M}O$_3$ crystals, and the main
reason why some of these compounds make it possible to combine
ferroelectricity (related to Bi's off-centering) and magnetism (associated
with the transition metals) at high temperatures.

Beyond its relatively small size, Bi$^{3+}$ presents an electronic
configuration (6$s^{2}p^{0}$) that allows for orbital rearrangements suitable
to form very directional bonds with neighboring oxygen atoms. Such Bi--O bonds
tend to result in a {\sl lone pair} on the non-bonding side, exactly as found
in BFO's $R3c$-G phase.\cite{ravindran06} This can be readily visualized in an
electron-localization-function (ELF)\cite{silvi94} analysis of the computed
electronic structure: As shown in Fig.~8(a), there is a distinct non-bonding
localization domain on the side of Bi that is opposite to the three
neighboring O atoms, which is the signature of a lone
pair.\cite{savin97,chesnut00} The occurrence of such a lone pair was discussed
at length by Ravindran {\sl et al}.\cite{ravindran06} on the basis of
first-principles calculations similar to ours; our results for BFO's $R3c$-G
phase (Figs.~8(a) and 9(a)) essentially reproduce their
study.\cite{fn:elf-values}

We computed the ELF maps for the other BFO phases found in this work. Figure~8
shows the results for two representative cases: $Cc$-C and $Pnma$-G. It is
immediate to note that a lone pair forms in the super-tetragonal $Cc$-C phase,
as might have been expected from Bi's large off-centering and the anisotropic
spatial distribution of its neighboring oxygens. The case of $Pnma$-G is quite
different, though: As shown in Fig.~8(c), in this phase the Bi cations have
four neighboring oxygens that form a rather regular BiO$_4$ tetrahedron. The
corresponding ELF plots show a very isotropic localization domain around
Bi. There is no clear lone-pair formation in this case; further, such a
localization domain is not typical of bonding electrons, as evidenced by the
slightly smaller ELF values along the directions of the Bi--O bonds. Hence, it
might be more appropriate to interpret this result as corresponding to a
semi-core-like case.  Interestingly, the partial density of states results
shown in Fig.~9 indicate that these three phases are very similar as regards
orbital occupation, even if they clearly differ in terms of Bi--O bonding and
lone-pair occurrence. Hence, our results illustrate Bi's electronic
flexibility and its ability to form different {\sl coordination complexes}
with the neighboring oxygens.

These chemical effects are clearly the driving force for the
structural transitions in BFO. Note that {\sl all} the BFO phases
discussed here, either ferroelectric or paraelectric, have an energy
that is lower than that of the cubic structure by more than
800~meV/f.u. (see Table~I). In contrast, the cubic and polar phases
differ by about 15~meV/f.u. in the case of prototype ferroelectric
BaTiO$_3$, where the Coulomb dipole-dipole interactions are known to
be the driving force for the FE instability.\cite{ghosez99} Noting
that BFO and BaTiO$_3$ are rather similar as regards the magnitude of
the dipole-dipole forces,\cite{fn:dipole} we can conclude that such an
enormous difference in the strength of the structural instabilities
must be associated with the dominant role of the Bi--O chemistry in
BFO. Then, the relative stability of BFO's low-energy phases is
probably determined by factors that involve smaller energy
differences, such as subtle competitions between different Bi--O
bonding mechanisms, the build-up of dipole-dipole interactions in the
FE phases, etc. Analyzing these issues in detail falls beyond the
scope of the present work. We hope our findings will stimulate further
theoretical studies of the chemical bond in these phases, so that the
factors controlling the occurrence of AFD and/or FE distortions
(especially the super-tetragonal ones) can be elucidated.

Let us conclude by noting that our results for BFO -- with most phases
being dominated by either AFD or FE distortions -- are clearly
reminiscent of the competition between AFD and FE instabilities that
is well-known to occur in many perovskite oxides. Such a competition
has been studied in detail in SrTiO$_3$,\cite{zhong95} and is one of
the factors responsible for the rich phase diagram of materials like
PbZr$_{1-x}$Ti$_{x}$O$_3$.\cite{kornev06} Interestingly, BFO is
peculiar inasmuch its FE soft mode in dominated by the A-site cation,
whereas ferroelectricity in SrTiO$_3$ is driven by the B-site
transition metal and PbZr$_{1-x}$Ti$_{x}$O$_3$ is an intermediate
case. Hence, BFO may constitute a new model system for the
investigation of competing-instability phenomena in perovskite oxides.

\section{Summary and conclusions}

We have used first-principles methods to perform a systematic search
for potentially-stable phases of multiferroic BiFeO$_3$. We worked
with a 40-atom super-cell (i.e., a 2$\times$2$\times$2 repetition of
the cubic perovskite cell) that is compatible with the atomic
distortions that are most common among transition-metal perovskite
oxides, namely, ferroelectric, anti-ferroelectric, and
anti-ferrodistortive. We obtained plenty of distinct low-energy phases
of the compound; here we have restricted the discussion to the most
stable ones. Many of the obtained minima present complex structural
distortions and very low symmetry (e.g., monoclinic $M_{A}$ and
$M_{C}$ space groups) while preserving a relatively small unit
cell. As far as we know, this is quite unique among perovskite oxides,
as the monoclinic structures reported so far are associated to complex
solid solutions (e.g., PbZr$_{1-x}$Ti$_{x}$O$_3$ or
PbZn$_{1/3}$Nb$_{2/3}$O$_3$-PbTiO$_3$), present large unit cells
(e.g., BiMnO$_3$ and BiScO$_3$), or are obtained under special
conditions (e.g., thin films subject to appropriate epitaxial strains
or bulk compounds under external electric fields). In contrast, our
study shows that bulk BiFeO$_3$ presents {\sl per se} a collection of
{\em simple} low-symmetry minima of the energy.

Our findings have a number of important implications for the research
on BiFeO$_3$ and related materials. Maybe the most general and
interesting one stems from the demonstration that BFO can form plenty
of (meta-)stable structural phases, which suggests that recent
puzzling observations -- ranging from possible structural transitions
at low temperatures\cite{lowT} to surface-specific atomic
structures\cite{marti-unp} and strain-induced new
phases\cite{zeches09,haumont09} -- may just be reflecting BFO's
intrinsic structural richness. Additionally, our results will provide
useful information to the experimental workers exploring the
possibility of obtaining large functional (piezoelectric,
magnetoelectric) effects in BiFeO$_3$ films grown on
strongly-compressive substrates: We have shown that there are plenty
of phases -- all with large polarizations and $c/a$ aspect ratios --
that can be realized in such conditions, including possibilities with
monoclinic and orthorhombic symmetries. Our results also provide new
insights concerning the relative importance of the various structural
distortions that can occur in BiFeO$_3$, stressing the key role that
the so-called {\sl secondary modes} play in determining the relative
stability of the observed phases.

Our work also has implications for theoretical studies of
BiFeO$_3$. First, we present a critical comparison of the various DFT schemes
most commonly employed to study BiFeO$_3$ and related compounds, and discuss
the existing difficulties to quantify the relative phase stability. Second, we
draw important conclusions as regards the effective modeling of structural
phase transitions in BiFeO$_3$, in connection with both Landau-type and
atomistic theories. Our analysis shows that BiFeO$_3$ is rather unique, and
that its modeling needs to address issues -- ranging from the work with
high-order Landau potentials to the accurate treatment of secondary
distortions -- that are unheard of in the work with {\sl classic} materials
such as BaTiO$_3$, PbZr$_{1-x}$Ti$_{x}$O$_3$, or even relaxor
ferroelectrics. Finally, our results provide quantitative evidence for the
dominant role that the Bi--O bond formation plays in BiFeO$_3$'s structural
instabilities. Further, our analysis suggests that some of the phases
discussed here do not exhibit the ``lone-pair mechanism'' usually invoked to
explain the Bi--O directional bonds in BiFeO$_3$. We take this as a new
illustration of Bi's ability to form diverse, competitive in energy, bonding
complexes with its neighboring oxygens.

In conclusion, we have used first-principles simulation methods to illustrate,
quantify, and analyze in some detail the structural richness of BiFeO$_3$, the
most relevant representative of the family of Bi-based transition-metal
perovskite oxides. Our simulations have revealed a variety of novel effects,
some of which have important implications for current experimental and
theoretical research on this material. We thus hope this work will help
clarify and further stimulate research on these ever surprising compounds.

Work supported by MICINN-Spain [Grants No. MAT2010-18113 and
No. CSD2007-00041, and {\em Ram\'on y Cajal} program (O.D.)] and by
CSIC's JAE-pre (O.E.G.V.)  and JAE-doc (J.C.W.)  programs. We used the
supercomputing facilities provided by RES and CESGA. We used the {\sc
  vesta}\cite{vesta} and {\sc Jmol}\cite{jmol} software for the
preparation of some figures, as well as the tools provided by the
Bilbao Crystallographic Server\cite{bilbao} and the {\sc isotropy}
group.\cite{isodisplace} Discussions with L.~Bellaiche, E.~Canadell,
G.~Catalan, L.~Chen, Z.~Chen, J.~Kreisel, J.F.~Scott, and M.~Stengel
are thankfully acknowledged.


\begin{thebibliography}{99}
\bibitem{fn:polarization} The reported ${\boldsymbol P}$ values were
  obtained as half of the polarization change between the
  ``$+{\boldsymbol P}$'' and ``$-{\boldsymbol P}$'' states. For the
  {\sl T} phases, special care was needed to identify an insulating
  path connecting such polar states, as we found that centrosymmetric
  structures with a large-$c/a$ cell tend to be metallic. Hence, our
  paths for the {\sl T} phases go through a small-$c/a$
  centrosymmetric structure.
\bibitem{fn:thermal} At finite temperatures, the relative phase
  stability will be determined by effects (e.g., vibrational entropy
  or thermal expansion) not considered here. Additionally, the
  transition temperatures may be strongly conditioned by the presence
  of competing instabilities\cite{zhong95} whose correct treatment may
  require a full statistical calculation.
\bibitem{fn:hydrostatic} We ran fixed-volume structural relaxations of the
  relevant phases in the range shown in Fig.~4.
\bibitem{fn:epitaxial} We ran epitaxially-constrained structural relaxations
  of the relevant phases in the range shown in Fig.~5. We forced the 40-atom
  cell of Fig.~1 to be square in the (001) plane.
\bibitem{fn:strain} Strictly speaking, the cell strain associated to
  the large $c/a$ ratios of our {\sl T} phases is a secondary order
  parameter. However, we recognize its obvious importance and do not
  consider it in this discussion.
\bibitem{fn:lower-order} As our discussion suggests, it might be
  possible to construct a lower-order Landau potential, and correctly
  account for all the low-symmetry minima, if additional degrees of
  freedom are explicitly considered in the theory. In the case of our
  simplified FE-only version of BFO, such additional degrees of freedom
  should be the cell strains.
\bibitem{fn:elf-values} As compared with the results of
  Ref.~\protect\onlinecite{ravindran06}, we obtained relatively small ELF
  values in the lone-pair region. Such differences are most likely related to
  some technicalities of the calculations, in particular to the use of
  different number of valence electrons.
\bibitem{fn:dipole} The magnitude of the dipole-dipole interactions is
  essentially controlled by the Born effective charges $Z^{*}$ (which
  quantify the electric polarization that appears as a consequence of
  an atomic displacement) and the electronic dielectric constant
  $\epsilon_{\infty}$ (which quantifies the system's ability to screen
  electrostatic interactions). All the atoms in BiFeO$_3$ and
  BaTiO$_3$ present large $Z^{*}$ values that exceed the nominal
  ionization charges by a factor of two at most; we obtained
  $\epsilon_{\infty}\approx$~7 for BaTiO$_3$ and
  $\epsilon_{\infty}\approx$~10 for BiFeO$_3$. Such deviations cannot
  account for the large difference in the energetics of the structural
  instabilities of these compounds.
\end{thebibliography}
\end{document}